\documentclass[traditabstract]{aa}
\usepackage{graphicx,subfigure,amsmath}
\usepackage{txfonts}
\usepackage{natbib}
\usepackage{ams}

\begin{document}

\title{Warm Debris Disks Candidates in Transiting Planets Systems}

\author{\'Alvaro Ribas \inst{1}
  \and Bruno Mer\'in \inst{1}
  \and David R. Ardila \inst{2}
  \and Herv\'e Bouy \inst{3}
}

\institute{Herschel Science Centre, European Space Astronomy Centre (ESA), 
  P.O. Box 78, 28691 Villanueva de la Ca\~nada (Madrid)
  \and NASA Herschel Science Center, California Institute of Technology, Mail Code 100-22, Pasadena, CA 91125, USA
  \and  Centro de Astrobiolog\'ia, INTA-CSIC, P.O. Box - Apdo. de correos 78, Villanueva de la Ca\~nada Madrid 28691, Spain}

\date{Received 21 October 2011 / Accepted 27 February 2012}

\abstract { We have bandmerged candidate transiting planetary
  systems (from the Kepler satellite) and confirmed transiting
  planetary systems (from the literature) with the recent Wide-field
  Infrared Survey Explorer (WISE) preliminary release catalog. We
  have found 13 stars showing infrared excesses at either 12\,$\mu$m and/or 22\,$\mu$m.  Without longer
  wavelength observations it is not possible to conclusively
  determine the nature of the excesses, although we argue that they
  are likely due to debris disks around the stars. If confirmed, our sample $\sim$doubles the number of currently known warm excess disks around old main sequence stars. The ratios
  between the measured fluxes and the stellar photospheres are
  generally larger than expected for Gyr-old stars, such as these
  planetary hosts.  Assuming temperature limits for the dust and
  emission from large dust particles, we derive estimates for the
  disk radii. These values are comparable to the planet's semi-major
  axis, suggesting that the planets may be stirring the
  planetesimals in the system.}

\maketitle

\section{INTRODUCTION}

Debris disks are the remnant of the planet formation process. The small detected amounts of dust are produced by collisions among, or evaporation of, planetesimals \citep[e.g.][]{Wyatt2008}. The dust generated by these processes is easier to detect than the planetesimals themselves.

Debris disks are not rare. According to \citet{Trilling2008}, $\sim$16\% of main-sequence FGK stars have debris disks. The infrared (IR) Spectral Energy Distributions (SEDs) of most stars with debris disks have peaks at 70-100\,$\mu$m, suggesting the presence of relatively cold dust (brightness temperatures $\sim$ 50 K).  

Overall, only $\sim$4\% of solar-type stars have $24\,\mu$m
excess flux, as seen with the {\it Spitzer Space Observatory} \citep{Trilling2008}. In most cases, the excess flux represents the Wien edge of longer-wavelength dust emission. However, a few systems (e.g. HD 23514, BD+20307, HD 69830, and
$\eta$ Corvi) posses real warm debris disks, and stand out above the envelope of 24$\mu$m
excess flux versus age \citep{Siegler2007}.

The presence of such large 24 $\mu$m excess at ages above the 100 Myr 
is problematic since neither the terrestrial-planet formation models by \citet{kenyon2005} nor
the steady state asteroid belt planetesimal grinding model by \citet{Wyatt2007b}, can explain
them. Their interpretation requires the warm dust to be very transient in nature. Such 
stochasticity could be due to dust input from a recent collision of planetesimals or
from a recent dynamical instability that essentially launched objects from the
outer disks into the inner disks \citep{Absil2006,Wyatt2008}. Planets at distances
at few AUs of their host star will have a strong impact in their asteroid belts
but so far very few host stars with both planets and warm excesses have been found (see  \citealt{Wyatt2008} for a review on the subject). 

In the spring of 2009, the {\it Kepler Mission} commenced high-precision photometry on over 150,000 stars, to determine the frequency and characteristics of transiting exoplanets. The target stars are determined by the Kepler Input Catalog
\citep[KIC,][]{KeplerInputCatalog,KICselection}. These are mostly main-sequence G type field dwarfs. On 2011 February, the Kepler Science Team released a list of 1235
candidate planetary systems orbiting around 997 host stars. The age of these stars is uncertain, although their low activity and low rotation velocities suggest Gyr ages \citep{Brown2011, Muirhead2011}.

The sample of planetary candidates is comprised mostly of objects with periods less than 1 yr. and a wide range of sizes. Analyses by~\citet{Morton2011} suggest
that 90 to 95\% of these candidates represent true planetary transits. This large sample of candidates provide us with a unique opportunity to study the incidence of debris disks around a homogeneous sample of planet-hosting candidate stars.

Within this context, on 2011 April the Science Team from Wide-field Infrared
Survey Explorer (WISE) released the Preliminary Release Catalog
(PRC)\footnote{http://wise2.ipac.caltech.edu/docs/release/prelim/}
covering 57\% of the sky and half of the Kepler field. WISE surveyed
the whole sky at 3.4, 4.6, 12, and 22 $\mu$m (hereafter, W1, W2, W3
and W4).

These two data releases allow us to explore the connection between
warm debris disks and transiting planets. In this paper we present the results of a search for warm debris disks, primarily around Kepler transiting planet candidates.  \citet{Krivov} have presented a similar study using pre-Kepler confirmed transiting systems, and in order to compare our technique with theirs we also include previously known transiting systems \footnote{http://var2.astro.cz/ETD/index.php} as described in \citet{exoplanetstransitdatabase}.

This first work aims at identifying bona-fide warm debris disk candidates that can afterwards be confirmed with deeper mid-IR imaging, either from the ground or from space.


\section{CANDIDATE SELECTION}

\subsection{Catalog Matching}
\begin{figure}
\includegraphics[width=\hsize]{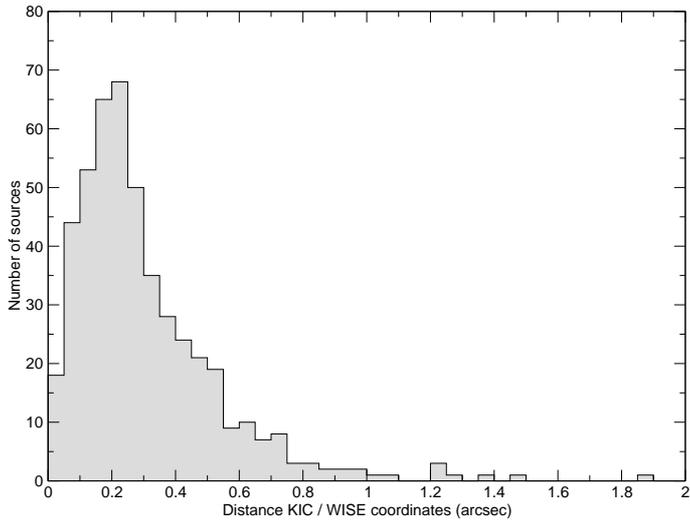}
\caption{Distribution of separations between the Kepler candidates and
  the WISE sources. A peak appears at 0.2\arcsec. We
  select a search radius of 1\arcsec in order not to loose potential
  candidates.}\label{distrib_r}
\end{figure}

We searched the WISE catalog for counterparts to the 997 Kepler
candidates planetary hosts and to the 131 known transiting planet host
stars by November 2011
\citep{exoplanetstransitdatabase}. Figure~\ref{distrib_r} shows the
distribution of separation between the KIC sources and their closest
match in the WISE catalogue within 10\arcsec. It appears to be roughly
normal at short distances with a peak around 0.2\arcsec and a standard
deviation of 0.2\arcsec, which is consistent with the astrometric
accuracy of the WISE PRC. Allowing a large search radius greatly
increases the probability of spurious matches, but limiting the search
radius to a small value would prevent us from detecting real high
proper motion objects. As a compromise, we limit the search to within
1\arcsec, corresponding to $\approx4\sigma$ with respect to the peak
in the separation distribution. With this process, 546 matches are
obtained (468 from the KIC and 78 from the known transiting planets
catalog).

Starting from a sample of 93 confirmed transiting hosts from exoplanets.org, \citet{Krivov} found 53 counterparts in the WISE sample. They do not describe the details of their matching protocol, but our detection fractions for non-Kepler objects are comparable to theirs, suggesting techniques with similar efficiencies. 

\begin{figure}
  \centering
  \includegraphics[width=\hsize]{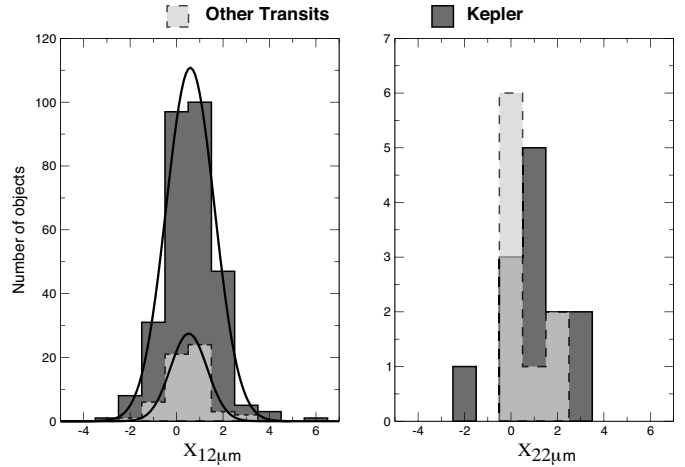}
  \caption{Histograms of $\chi_\lambda \equiv (F_{\rm WISE}-F_{\rm
      phot}) / \sigma_{tot}$, for the 12 and 22\,$\mu$m bands, for sources
    with S/N$>$3 in the respective bands.The KIC (350 objects in the
    W3 band and 31 in the W4 band) and known planets samples (61 objects
    in the W3 band and 21 in the W4 band) are shown in dark and light grey
    respectively. A gaussian distribution is fitted for the 12\,$\mu$m
    band,($x_0=0.5,\sigma=1$).}
  \label{fig:hist}
\end{figure}

\subsection{SED Fitting}

\begin{table}
  \begin{center}
    \begin{tabular}{l c c l}
      \hline\rule{0mm}{3mm} Name & ${\chi_{12}}$  & ${\chi_{22}}$ & Quality \vspace{1mm}\\
      \hline
      \vspace{-2.5mm}\\
      KIC~2853093 & 6.1 & 2.2 & 1 - \checkmark\\
      KIC~9703198 & 4.2 & \ldots & 1 - \checkmark\\
      KIC~6422367 & 4.0 & \ldots & 1 - \checkmark\\
      KIC~4918309 & 2.2 & \ldots & 1 - \checkmark\\
      KIC~6924203 & 2.2 & \ldots & 1 - \checkmark\\
      KIC~9008220 & 2.2 & \ldots & 1 - \checkmark\\
      KIC~10526549 & 2.1 & \ldots & 1 - \checkmark\\
      WASP-46 & 3.2 & \ldots & 2 - Nearby sources in W3\\
      KIC~3732821 & 2.6 & \ldots & 2 - Nearby source in W3\\
      KIC~4545187 & 2.6 & \ldots & 2 - Slight offset in W3\\
      KIC~6665695 & 0.8 & 2.2  & 2 - Nearby sources in W3 and W4\\
      KIC~3547091 & 2.2 & \ldots & 2 - Slight offset in W3\\
      KIC~8414716 & 2.2 & \ldots & 2 - Slight offset in W3\\
      KIC~11288505 & 3.8 & \ldots & 3 - Extended in W3 \\
      KIC~6266741 & 3.4 & \ldots & 3 - Extended in W3\\
      KIC~9513865 & 3.3 & 2.5 & 3 - Extended in W3 and W4\\
      KIC~5184584 & 2.2 & 3.1 & 3 - Offset in W3; Extended in W4 \\
      KIC~7031517 & 3.0 & \ldots &  3 - Extended in W3\\
      KIC~2309719 & 0.6 & 2.8 & 3 - Extended in W4\\
      KIC~9007866 & \ldots & 2.8 & 3 - Offset in W3\\
      KIC~10963242 & 2.5 & \ldots & 3 - Offset in W3\\
      CoRoT-14 & 2.5 & \ldots & 3 - Artifact in W3\\
      KIC~3833007 & \ldots & 2.4 & 3 - Offset in W3; Extended in W4\\
      KIC~4544670 & 2.4 & \ldots & 3- Extended in W3\\
      KIC~7663405 & 2.3 & \ldots & 3 - Offset in W3 \\
      KIC~9214942 & 0.9 & 2.3 & 3 - Offset in W3; Extended in W4\\
      KIC~3642741 & 2.2 & \ldots & 3 - Extended in W3\\
      KIC~9757613 & 1.9 & 2.1 & 3 - Extended in W4\\
      KIC~11446443 & 0.8 & 2.1 & 3 - Offset in W4; TrES-2\\
      HAT-P-5 & 0.6 & 2.1 & 3 - Offset in W4\\
      KIC~6692833 & 1.9 & 2.0 &  3 - Extended in W3 and W4\\
      KIC~9391506 & \ldots & 2.0 & 3 - Offset in W3 and W4\\
      CoRoT-8 & -1.4 & 2.0 & 3 - Offset in W4\\
      \hline\rule{0mm}{2.0mm}
    \end{tabular}
  \end{center}
\caption{Sample of the 33 candidates after performing the
  cut above $\chi_{12} \ge 2$ or $\chi_{22} \ge 2$. From the SED fitting, typical uncertainties are 0.2 and 0.02 at 12 and 22 $\mu$m, respectively. Where marked as ellipsis (\ldots), the WISE measurements are upper limits. For each candidate we have inspected the WISE images and established a quality parameter as: (1) Good, clear WISE detection;  (2) Possible WISE detection; (3) Rejected. The reasons for rejection are clear image artifacts, offset in the photocenter of the image as compared to shorter wavelengths, and "extended" sources. "Extended" sources are ones in which the images show deformed PSFs, either due to a real extended object, to confusion with nearby objects, or to confusion with noise.}\label{tab:candidates}
\end{table}

In order to obtain predictions for the photospheric values, we compared the observed SED in the optical and near IR with stellar models, using the VO SED Analyzer
\citep[VOSA,][]{VOSA}. VOSA is a tool designed to query VO-compliant
theoretical model spectra, calculate the synthetic photometry in any
given filter set, and then perform $\chi^2$ minimization to determine
a best fit to the data. We chose to compare the observed SEDs with the
grid of Kurucz ATLAS9 models \citep{Castelli1997} using effective
temperatures in the range 3500 $\leq T_{\rm eff} \leq$\,9250~K, covering
exactly the range of estimated effective temperature in our targets. For 
all sources, the broadband photometry given as input to VOSA included at least 
$J$, $H$ and $Ks$ from 2MASS~\citep{2MASS}, and the associated {\it WISE} photometry. In the
case of the Kepler sources, we added the $u,g,r,i,z$ photometry.  For the other sources, we added SDSS\citep{SDSS} and/or Tycho \citep{Tycho} photometry for sources within a
radius of 1\arcsec. 

VOSA use as inputs all the photometry available plus an estimate
of the interstellar extinction for each source to perform an optimal
fit of the photometry to a stellar atmosphere model. We used as input the value of the extinction provided in the KIC, with the extinction law
by \citet{Cardelli1989}. For the non-Kepler sources, we have assumed $A_V$=0.0\,mag,  a good approximation to the actual extinctions, as found by Krivov et al. (2011) in the objects in common. 

The fit of the optical and near-IR photospheres  is degenerated
in the ($A_V$, T$_{\rm eff}$) plane. Higher effective temperature stars with
higher extinction effectively have a similar observed SED than lower temperature
stellar photospheres with less extinction.  However, by allowing the extinction to vary as a free parameter, we find typical ranges of $\Delta T_{\rm eff} <$ 200 K and $\Delta A_V < 0.5$  mags, where  the SED fits result in similar $\chi^2$ values. The best-fit effective temperatures and $A_V$ values computed by VOSA in this way are in excellent agreement with those provided by the KIC, with a scatter smaller than 10\%. This confirms that the stellar SED fits are sound and good enough for the purposes of this analysis. 

To identify the excess sources, we selected those objects not flagged as
extended in the WISE catalog, with signal-to-noise ratios (S/N) $\ge$ 3 at W3, and for
which $\chi_\lambda \equiv (F_{\rm WISE}-F_{\rm phot}) / \sigma_{tot} \ge 2 $. Here
$F_{\rm WISE}$ is the detected, de-redenned flux at W3 or W4 band, $F_{\rm phot}$
is the corresponding synthetic flux value obtained from photospheric modeling,
and $\sigma_{tot}$ represents the total uncertainty computed as
$\sigma_{tot}=\sqrt{\sigma^{2}_{\rm obs} + \sigma^{2}_{\rm cal}}$, being
$\sigma_{\rm obs}$ the measurement uncertainty in the corresponding
band, and $\sigma_{\rm cal}$ the absolute calibration uncertainty of
the WISE instrument (2.4, 2.8, 4.5 and 5.7\% for W1, W2, W3 and W4
bands respectively as indicated in the Explanatory Supplement to the
WISE Preliminary Data Release
Products\footnote{http://wise2.ipac.caltech.edu/docs/release/prelim/expsup/}).
The error associated to the fit done by VOSA was estimated by
comparing the results obtained using the 5 best fits. The variations
of the photospheric synthetic fluxes in the WISE bands were found to
be negligible and were therefore not considered. This result in consistent with the negligible values for such uncertainty found by Krivov et al. (2011). The $\chi_\lambda$ distributions for W3 and
W4 are shown in figure~\ref{fig:hist}. We found 33 candidates with $\chi_\lambda>2$ in one or both bands. 

Given the steep decline of the interstellar extinction law towards the
mid-infrared, variations in up to 1 mag in $A_V$ will only raise the
WISE fluxes by 6, 5, 7 and 3\%, at the W1, W2, W3 and W4 bands,
respectively, and therefore cannot dominate the final uncertainty
budget. In fact, the intrinsic uncertainties in the $A_V$ and T$_{\rm
  eff}$ from the SED fitting allow variations of $\chi_\lambda$
smaller than 0.2 and 0.02 at W3 and W4 bands, respectively.  In other
words, no reasonable combination of T$_{\rm eff}$ and $A_V$ will make
these excesses disappear.

Figure \ref{fig:hist} shows the $\chi_\lambda$ distributions. Both are
centred around 0.5, and have a standard deviation of 1. We choose to
set a threshold of $\chi_\lambda \ge 2$ , corresponding to
1.5\,$\sigma$, or a 87\,\% confidence level, to conservatively define
good excesses in both bands. The analysis of the origin of this shift
in the excess significance distribution would be premature given the
early status of the release of the WISE PRC.  It must be noted in any
case that the value of our threshold to define an excess has been set
arbitrarily to allow us to identify bona-fide excess targets which can
then be confirmed by longer wavelength and/or deeper mid-IR
ground-based observations. The process will certainly be repeated once
the final WISE catalog will be released with the latest updates on
calibration and covering the whole Kepler field.

The analysis also identifies excesses at the W1 and W2 bands in KIC
2853093 and KIC 3547091. The interpretation of these as very hot dust
emission is problematic, given that Spitzer did not detect large
populations of very hot debris disks at the analogous bands (IRAC1 and
IRAC2). We do not consider these excess further in our study. We will
revisit the issue once the final WISE catalog, with the final
photometric reprocessing, is made public.

\subsection{Image Inspection}

\begin{figure*}
  \centering
    \subfigure{\includegraphics[width=.305\hsize]{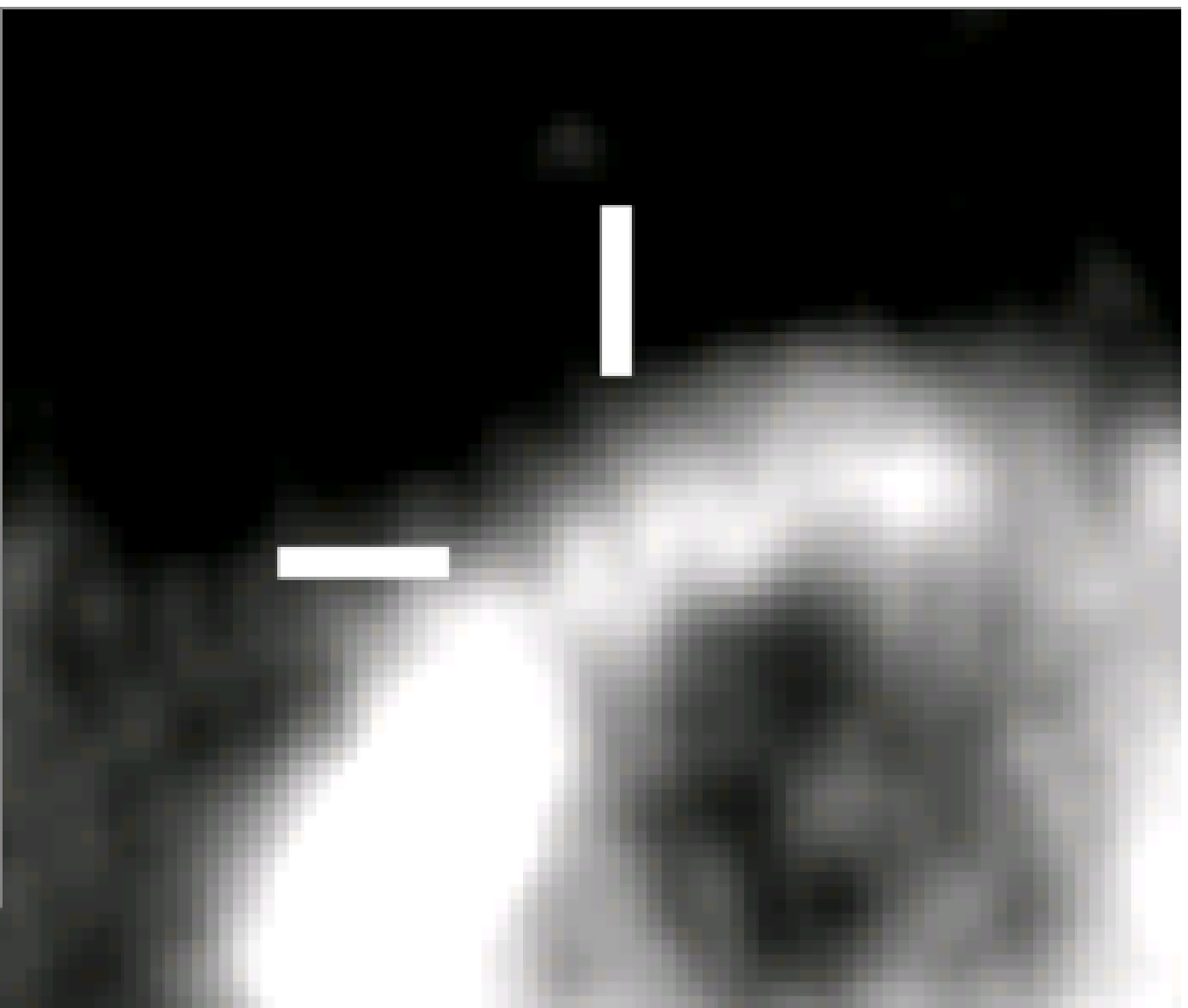}}
    \subfigure{\includegraphics[width=.31\hsize]{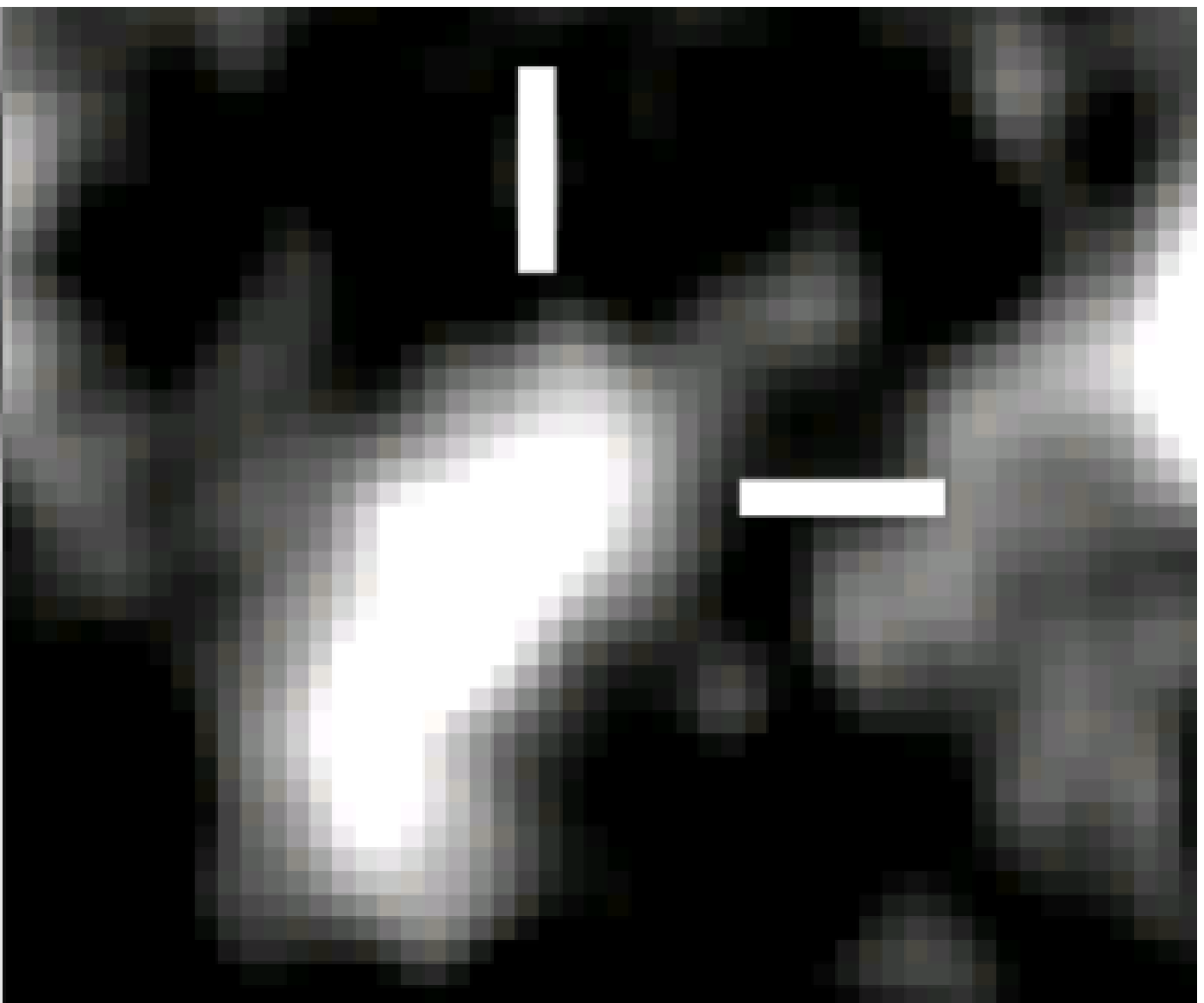}}
    \subfigure{\includegraphics[width=.3\hsize]{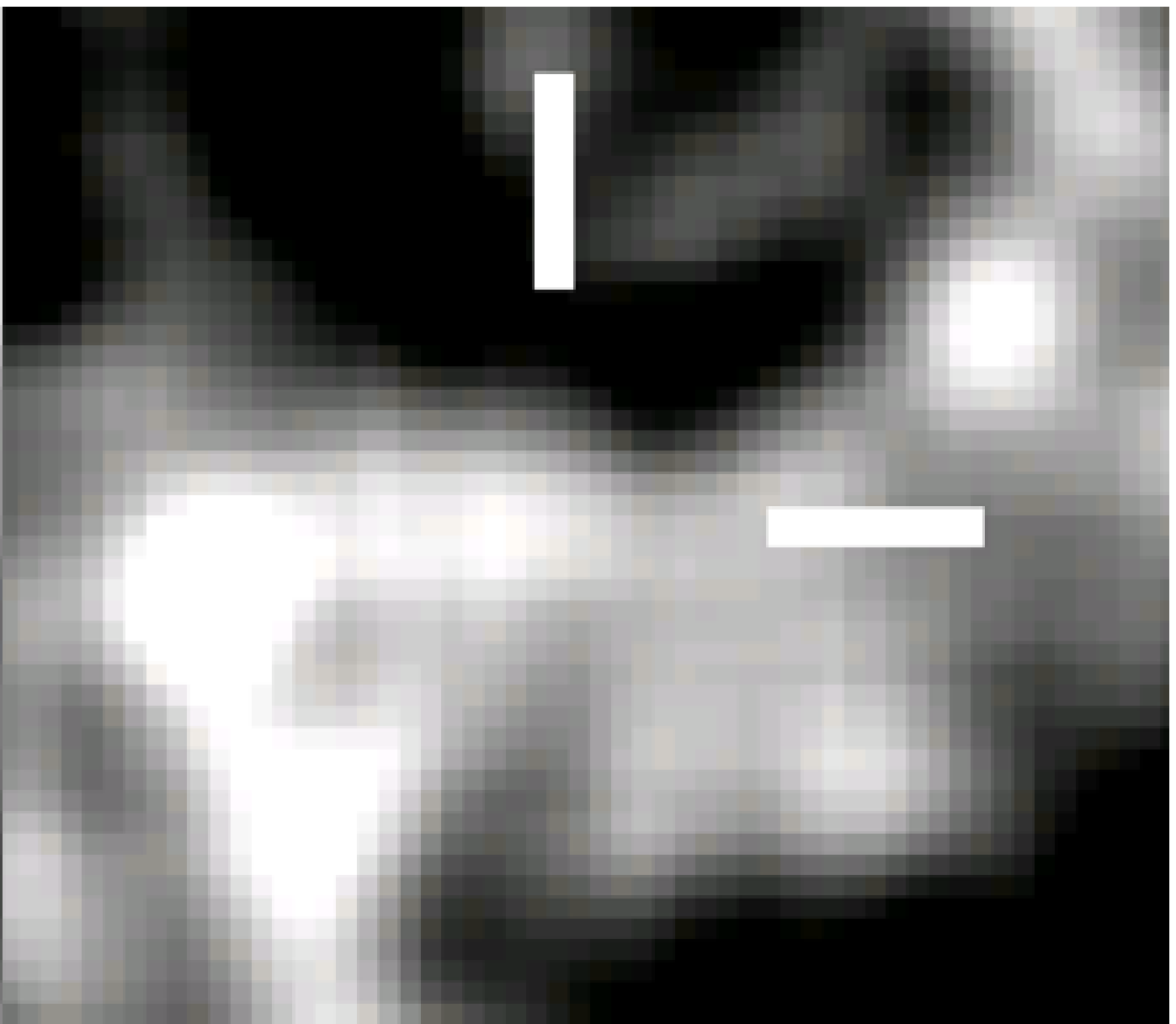}}\\
    \subfigure{\includegraphics[width=.3\hsize]{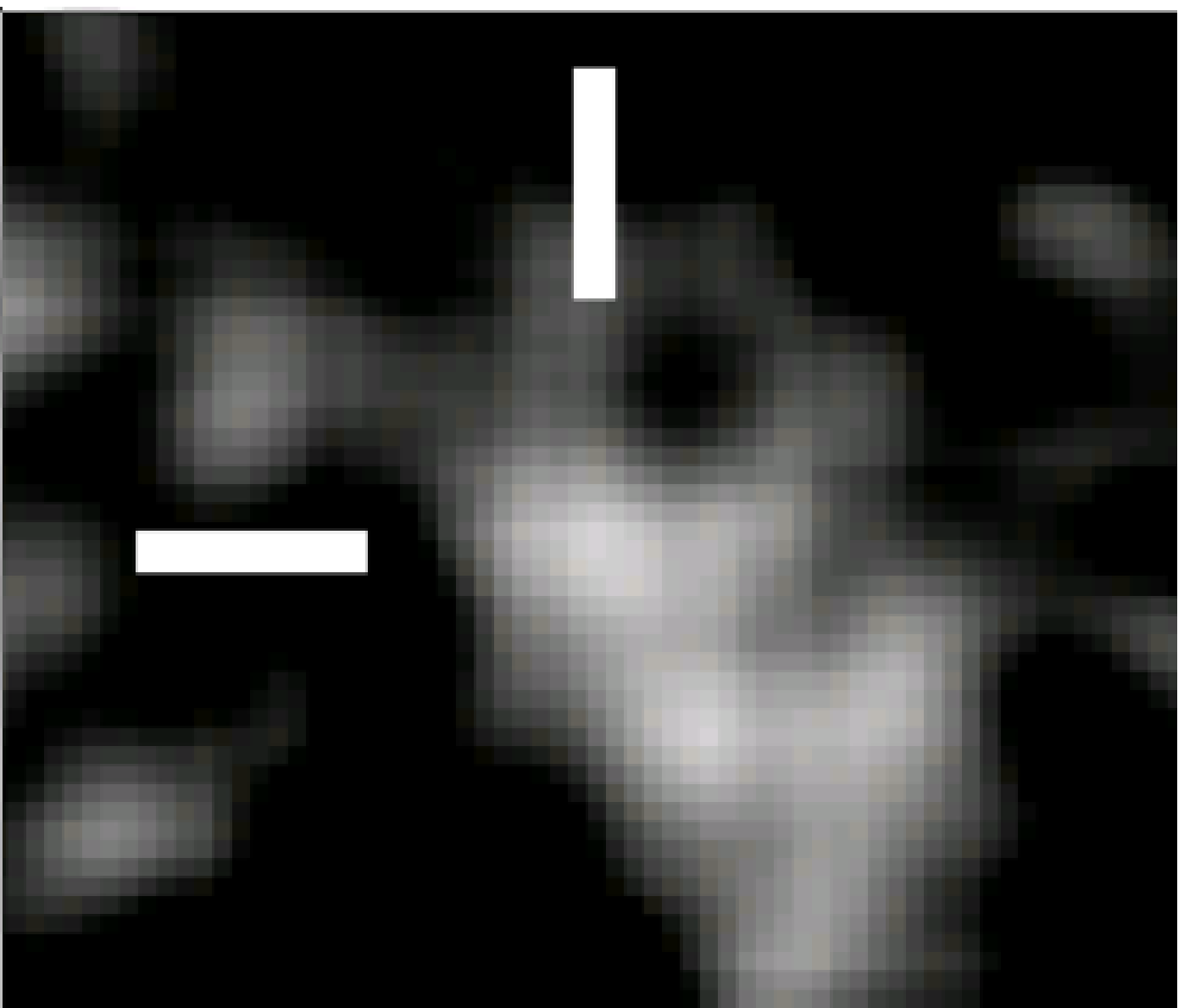}}
    \subfigure{\includegraphics[width=.3\hsize]{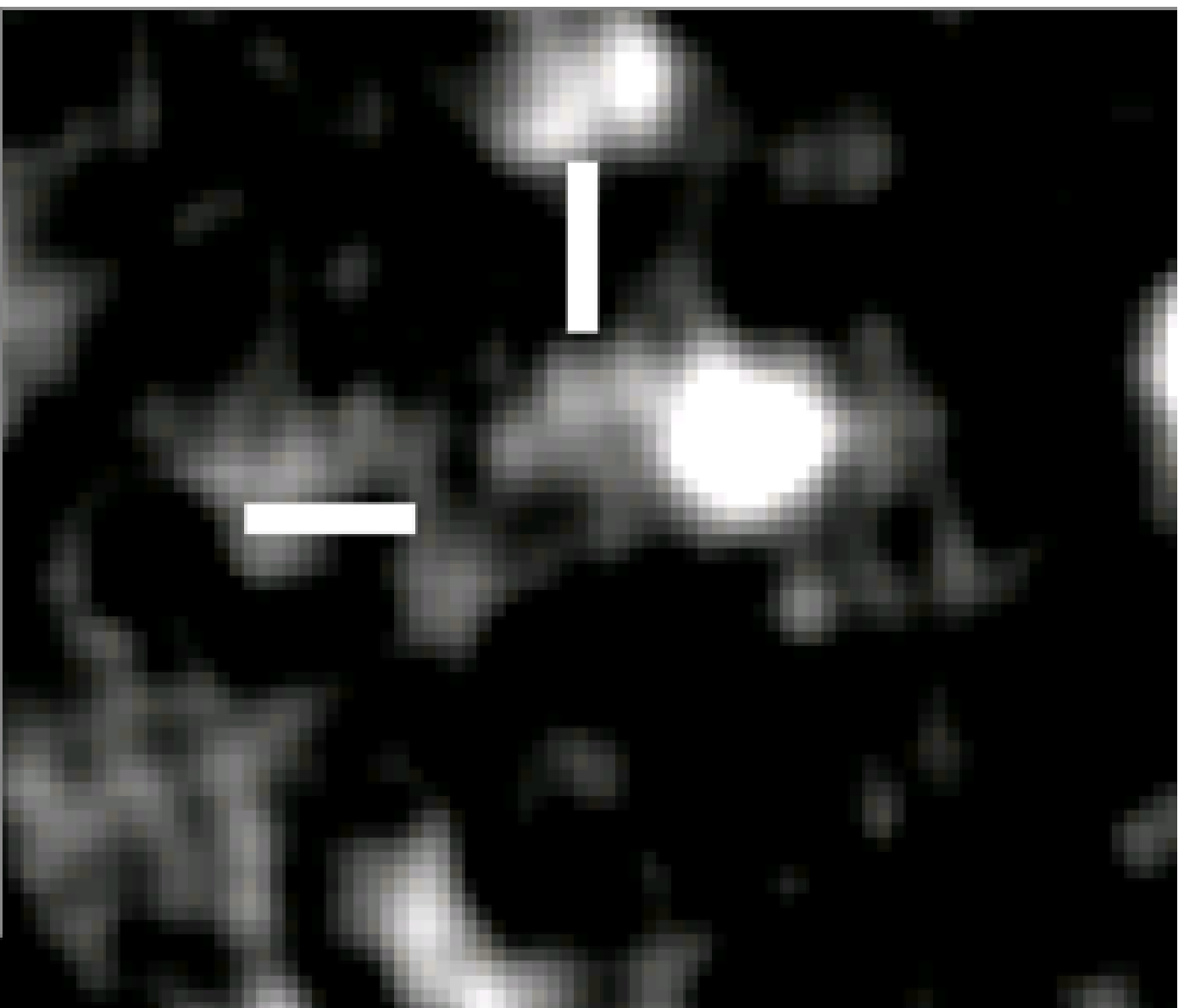}}
    \subfigure{\includegraphics[width=.3\hsize]{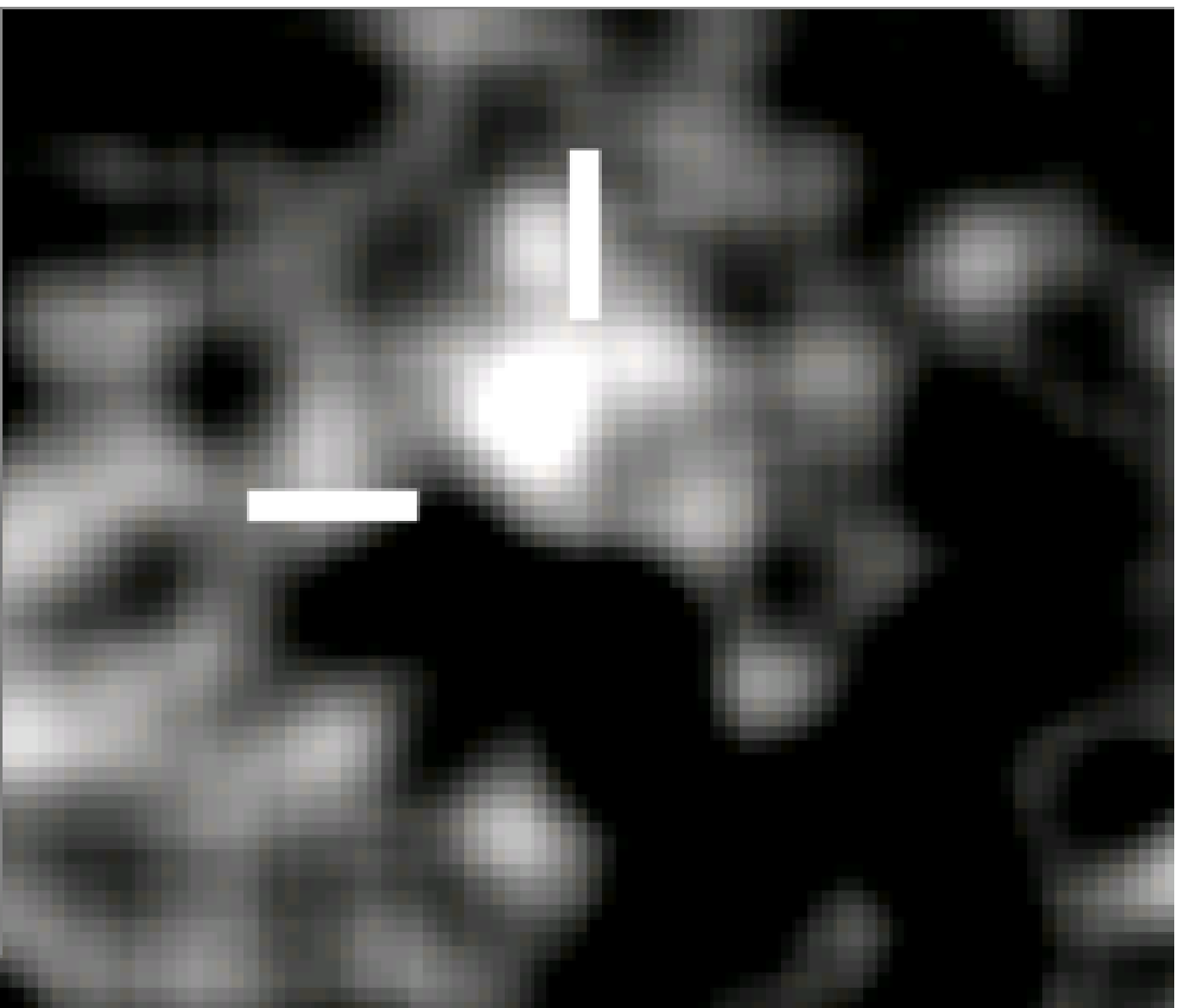}}    

\caption{Here and in the next two figures, North is up and East is left.  Each image is 1.7' x 1.9'. The thick segments indicate the nominal position of the target in the WISE catalog. The color scale goes from zero to 3$\sigma$, where the sigma is calculated as the standard deviation of the pixels in the image. Examples of different problems in the images: (Top Left) Artifact in the W3 band for CoRoT-14; (Top Middle) Extended PSF in W3 for KIC~11288505; (Top Right)  Extended PSF possibly due to source confusion in W4 for KIC~6692833; (Bottom Left) Extended PSF possibly due to low S/N in W4 for KIC~2309719; (Bottom Center) Photocenter offset in W3 and (Bottom Right) W4 for KIC~9007866.}\label{fig:problems}
\end{figure*}

\begin{figure*}
  \centering
    \subfigure{\includegraphics[width=.23\hsize]{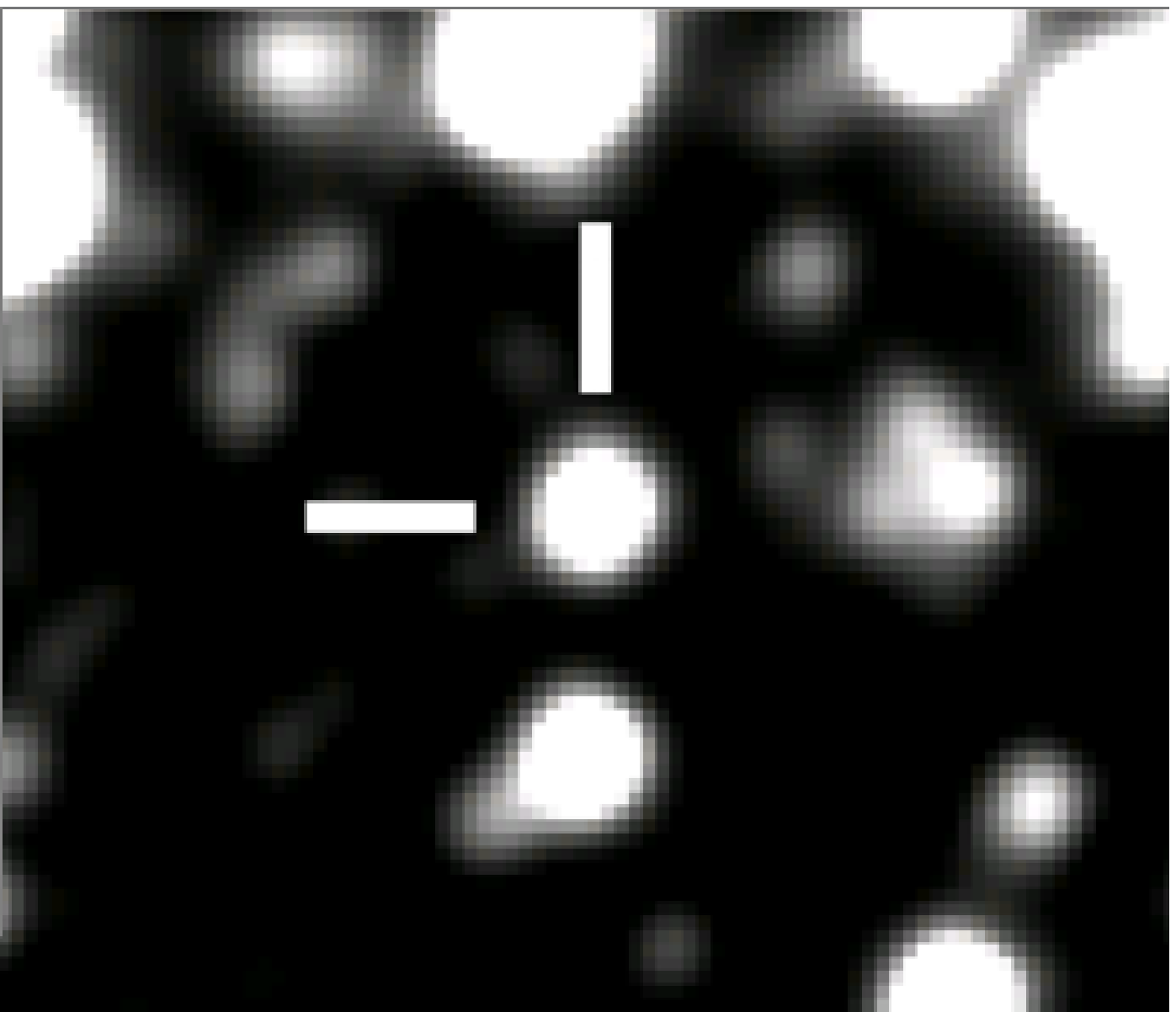}}
    \subfigure{\includegraphics[width=.23\hsize]{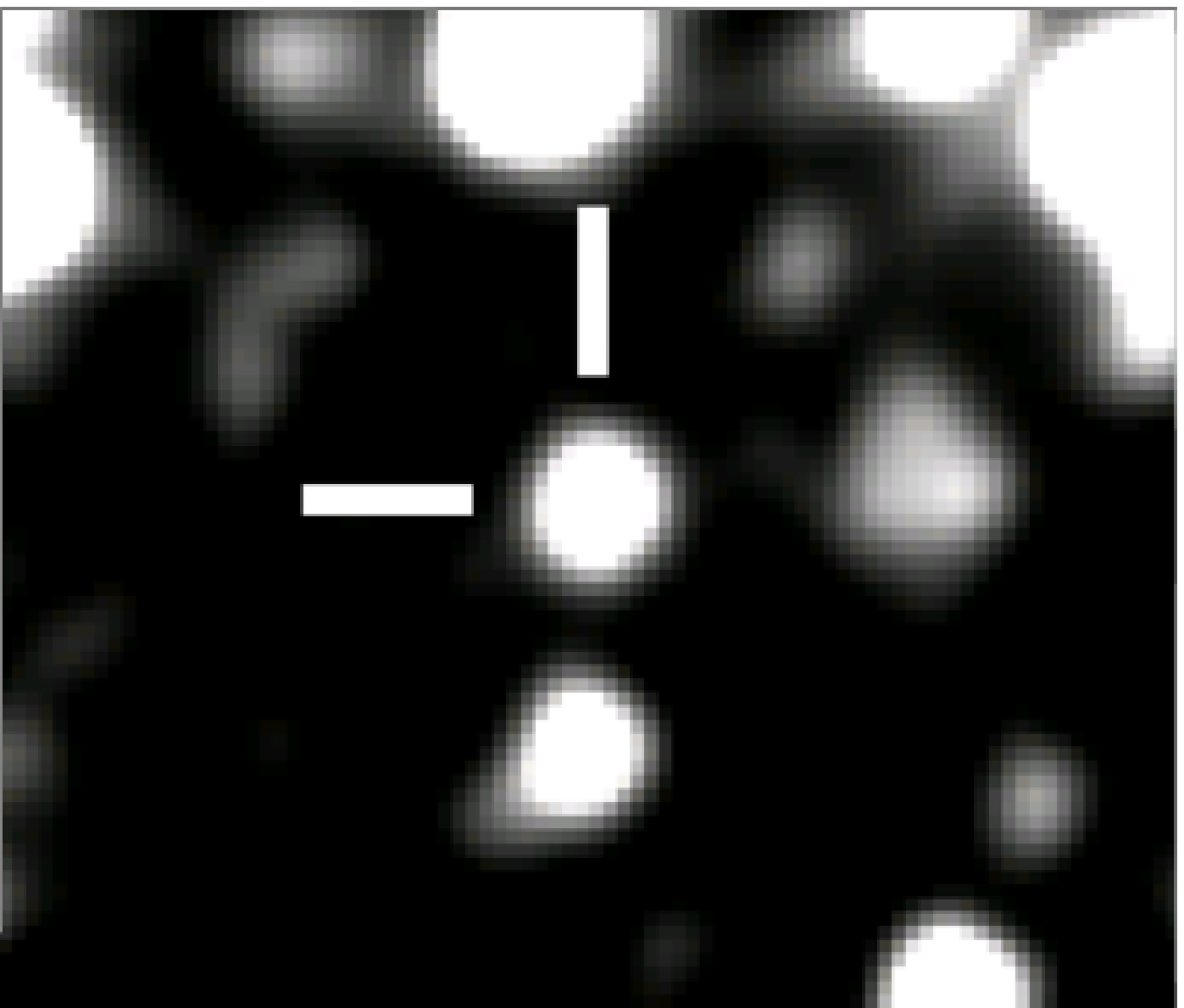}}
    \subfigure{\includegraphics[width=.23\hsize]{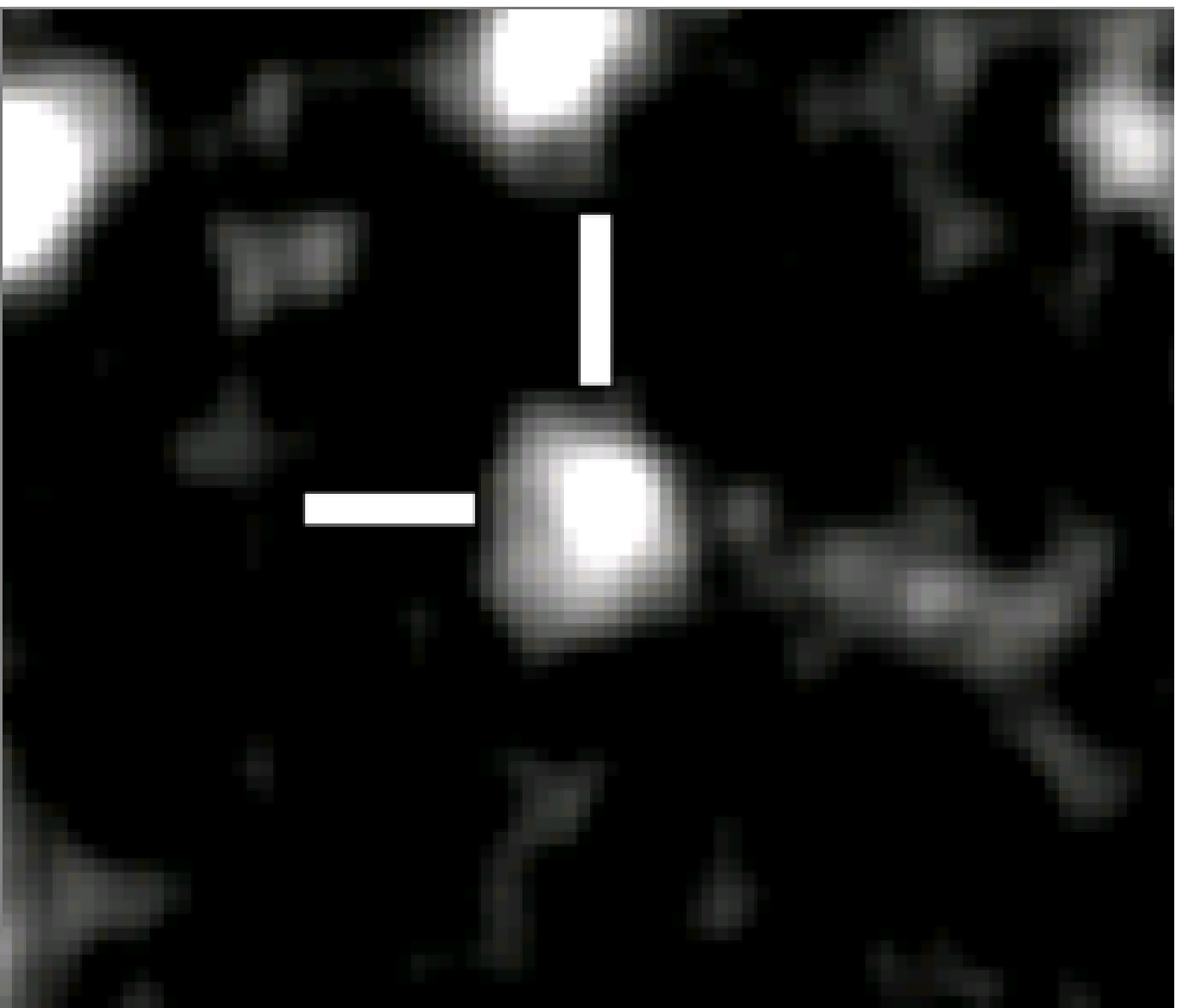}}
    \subfigure{\includegraphics[width=.23\hsize]{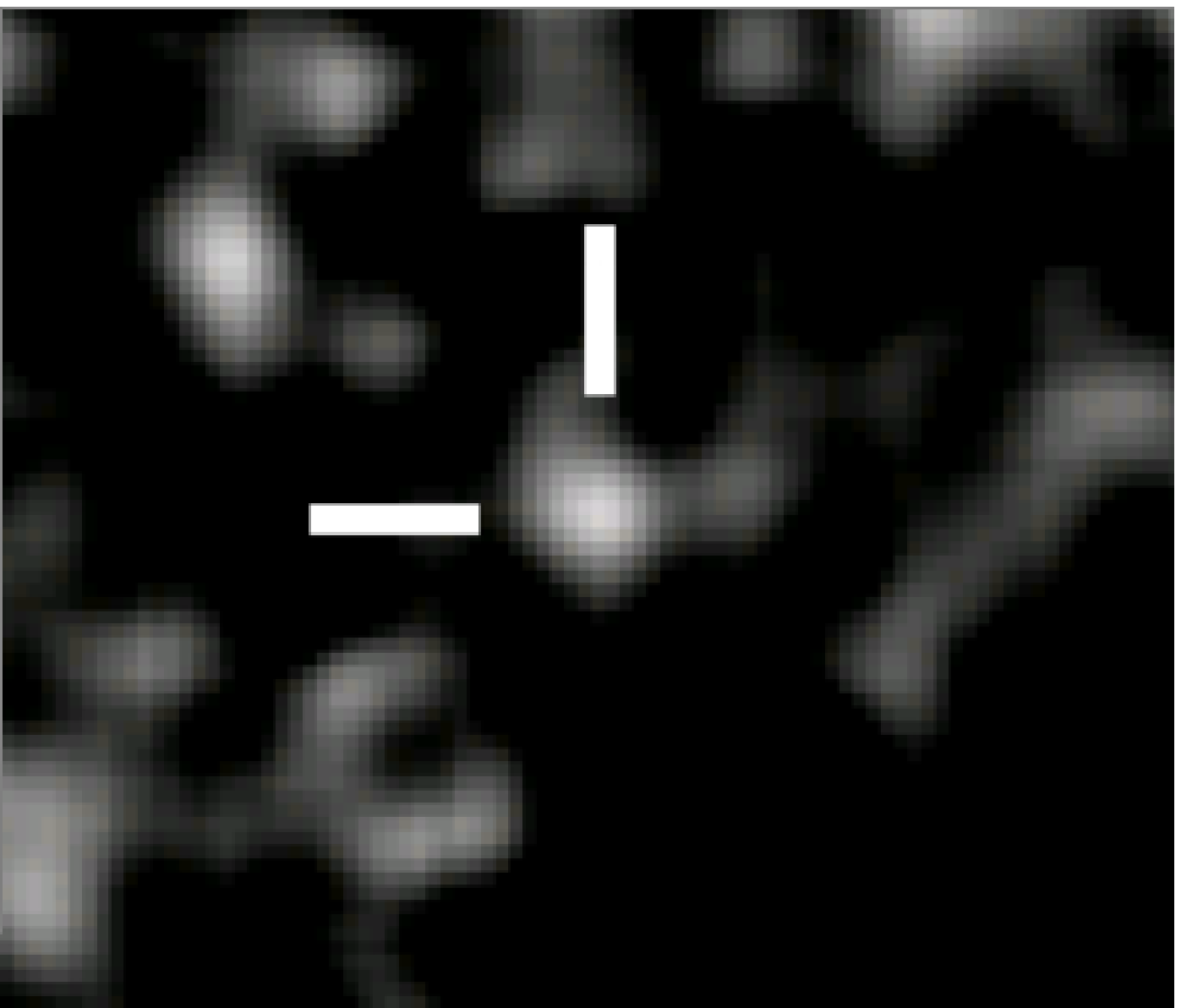}}  
\caption{KIC~2853093: A clear detection of a single isolated PSF source with excess at W3 and W4. The four bands (from left to right W1, W2, W3, and W4) are shown.There is no source confusion and the centroid position is maintained in every image. Most other point source features in the image have been checked to match other 2MASS point sources.}\label{fig:clear}
\end{figure*}

Table~\ref{tab:candidates} lists the 33 host stars with candidate
  excesses. We examined the WISE images at the bands in which the excesses were present and assigned a quality flag to each candidate as follows: (1) Good, relatively isolated PSF in the band(s) at which the excess is detected; (2) Slight offsets with respect to shorter wavelengths, or slight PSF extension, in the band(s) at which the excess is detected; (3) Rejected: fundamental problems with the band(s) at which the excess is detected. We reject targets based on three issues (see Figure \ref{fig:problems}):
\begin{itemize}
\item Artifacts: This affects only CoRoT-14 (see Fig. \ref{fig:problems}).
\item Offsets: The position of the photocenter changes at the band where excess is found compared to its position at W1 or W2 bands. 
\item "Extended" PSFs: The photocenter at the band where the excess is found does not appear as a single PSF. This may be because the source is extended, because additional sources are present nearby at the same flux level, or because the source is faint enough that significant noise can contribute to the detection. 
\end{itemize}
In total, we identify 7 objects with quality flag (1) and 6 objects with quality flag (2). Of both sets, twelve objects show IR excesses with $\chi_{12} > 2$ at W3, and two sources have $\chi_{22} > 2$ at W4 (See Figure \ref{fig:clear}). 

For  KIC~9008220, KIC~4918309, and KIC~3547091, the WISE catalog flags the photometry in the W1 and W2 bands as affected by halo features, the outer wings of the PSF around nearby bright stars. Likewise the same bands in KIC~8414714 are flagged as
affected by diffraction spikes. None of these issues is apparent in
the actual images nor in the SEDs (see below). For KIC~9703198, KIC~8414716, and
KIC~2853093 the W4 band is flagged as having 1\% of saturated
pixels. This does not affect that band's fluxes and will not affect
our analysis.

\begin{figure*}
  \centering
    \subfigure{\includegraphics[width=.3\hsize]{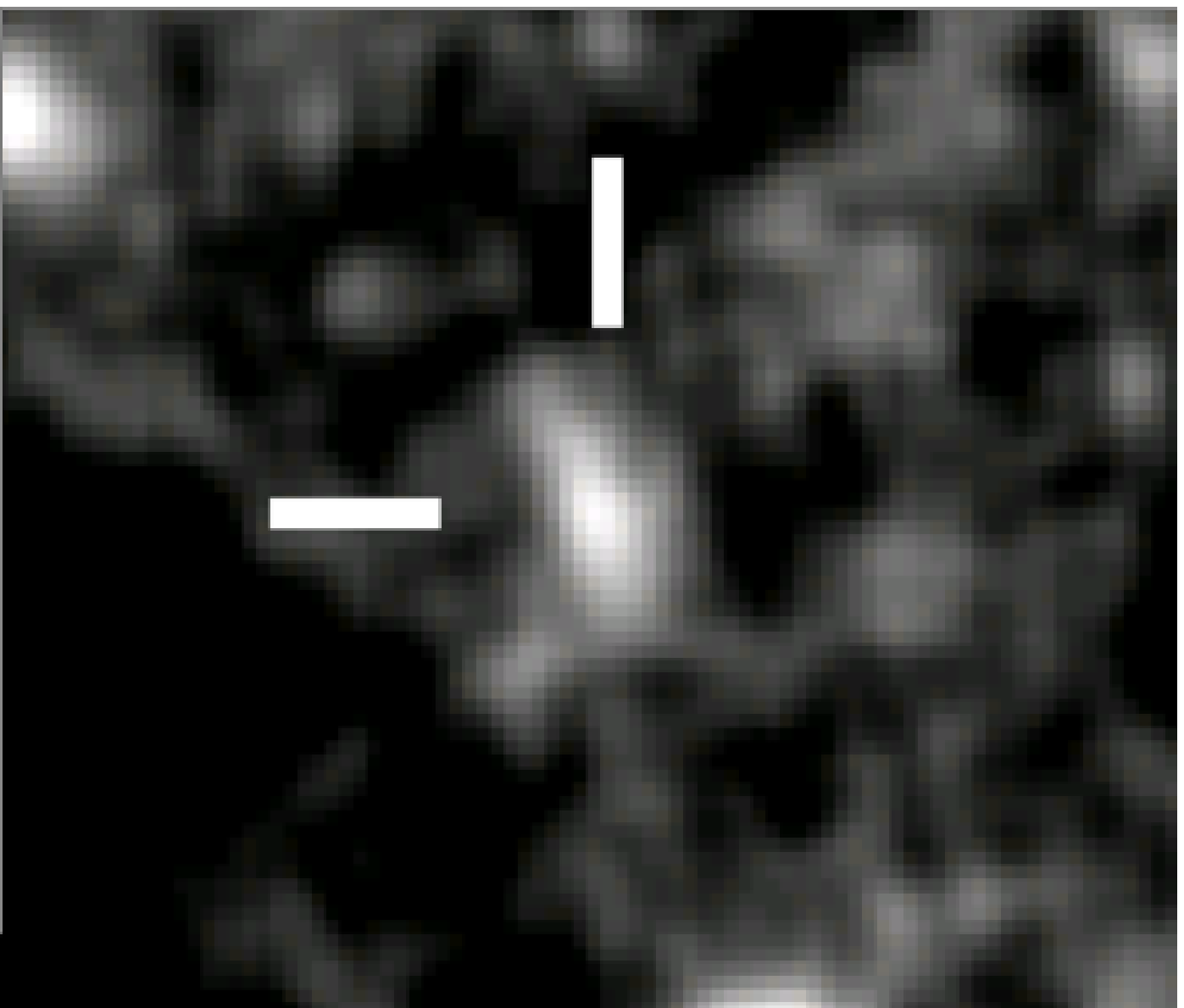}}\,\,\,\,\,
     \subfigure{\includegraphics[width=.3\hsize]{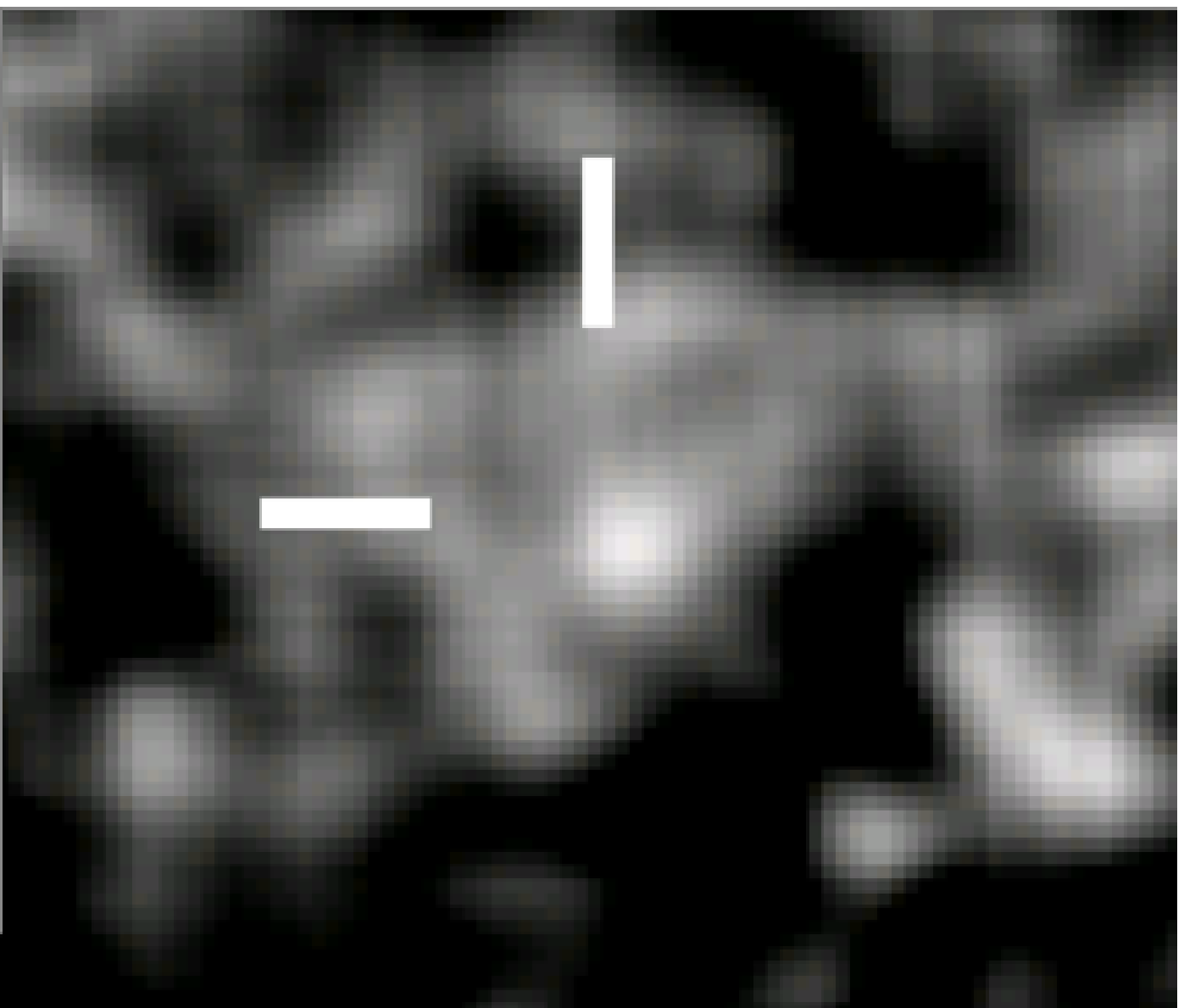}} \\
    \subfigure{\includegraphics[width=.3\hsize]{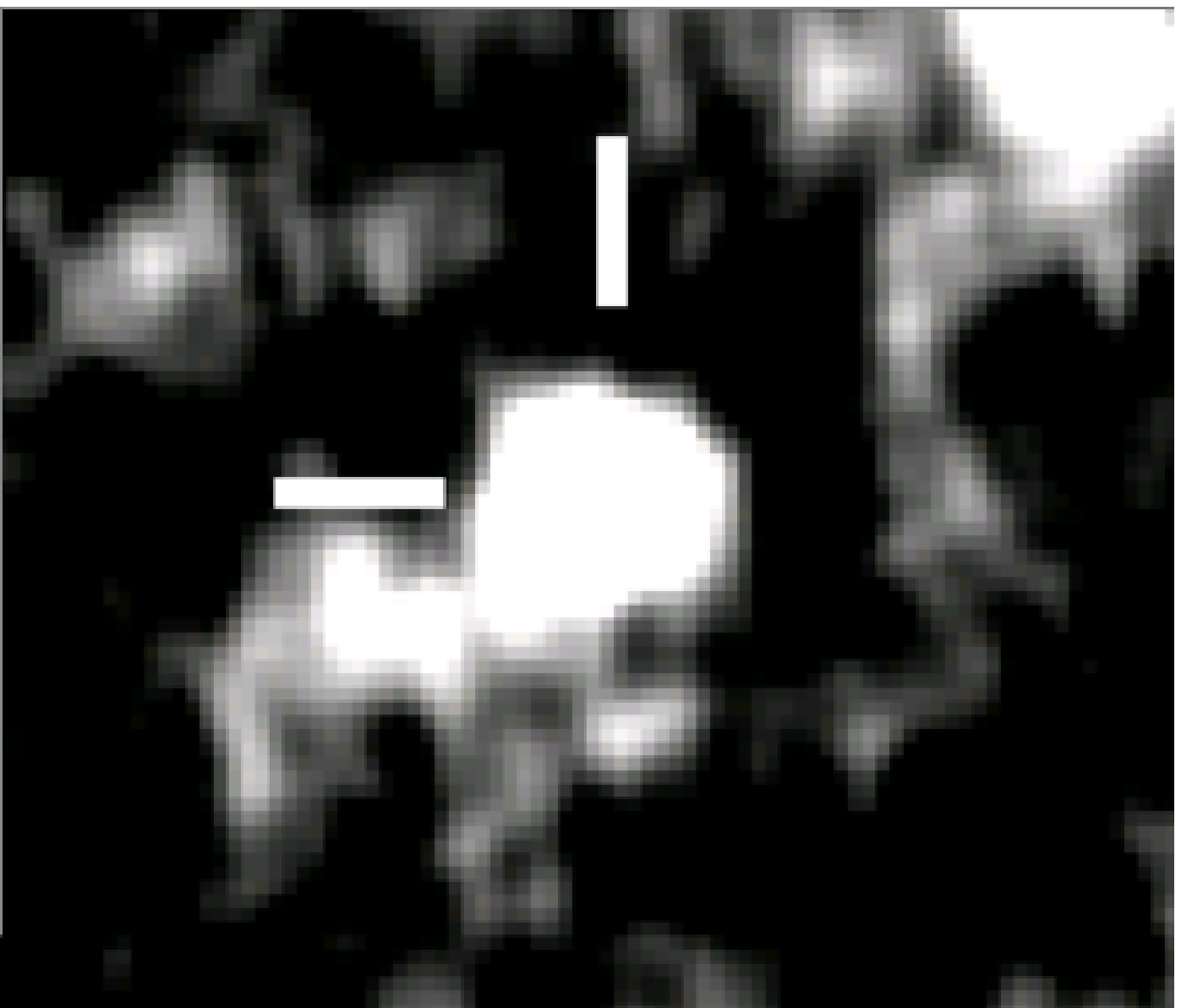}}\,\,\,\,\,  
     \subfigure{\includegraphics[width=.3\hsize]{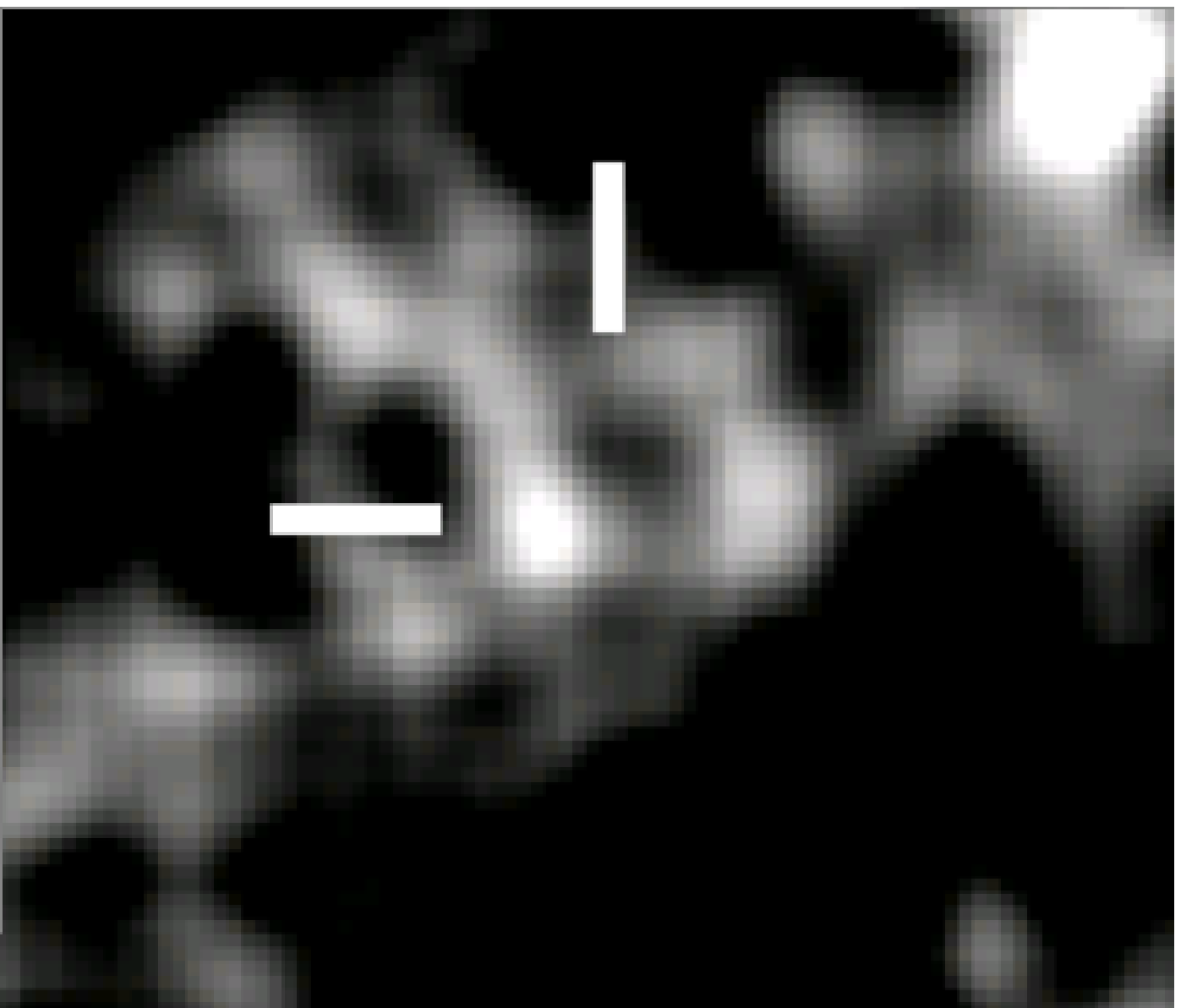}} \\ 
     \subfigure{\includegraphics[width=.3\hsize]{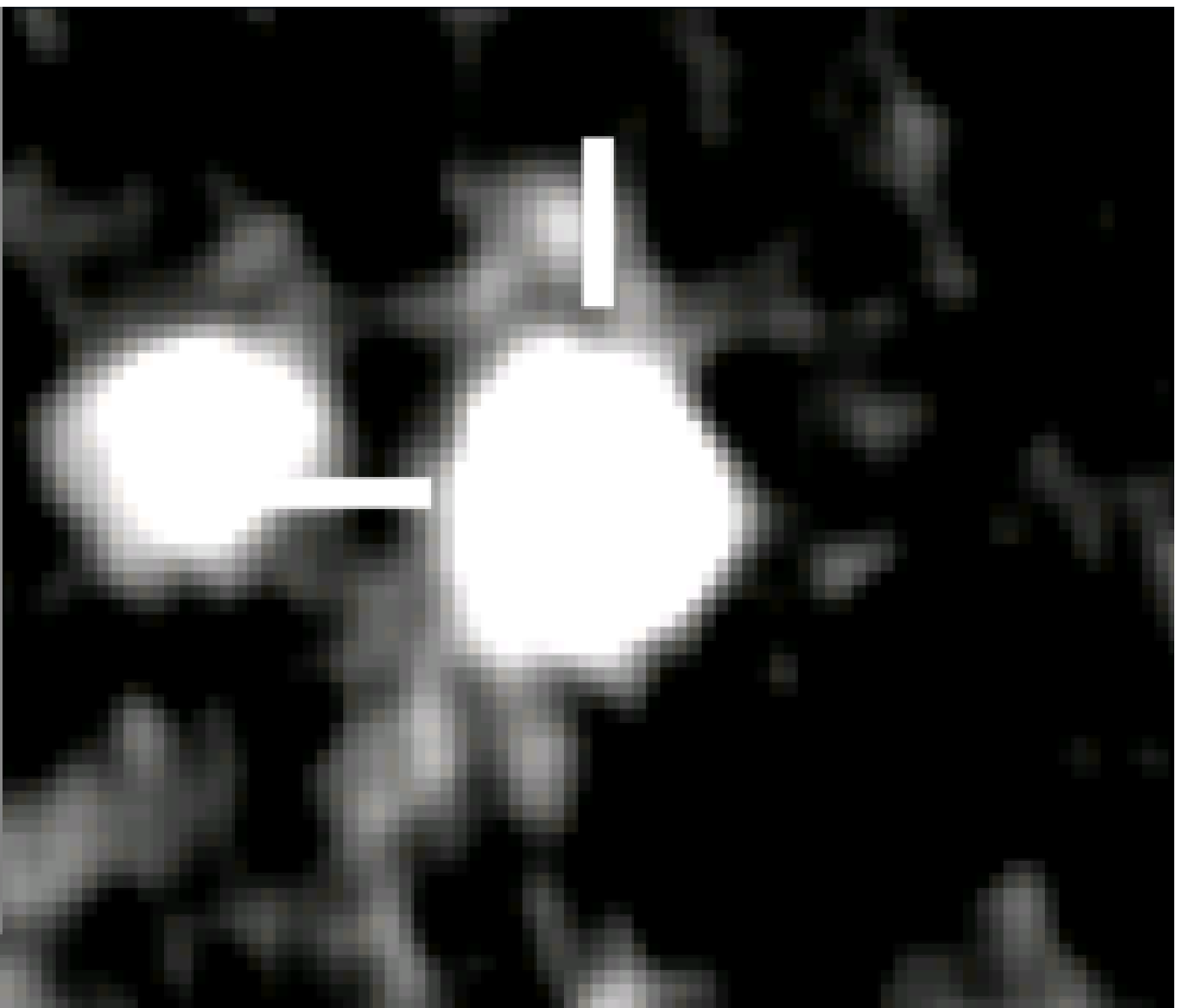}}\,\,\,\,\,
     \subfigure{\includegraphics[width=.3\hsize]{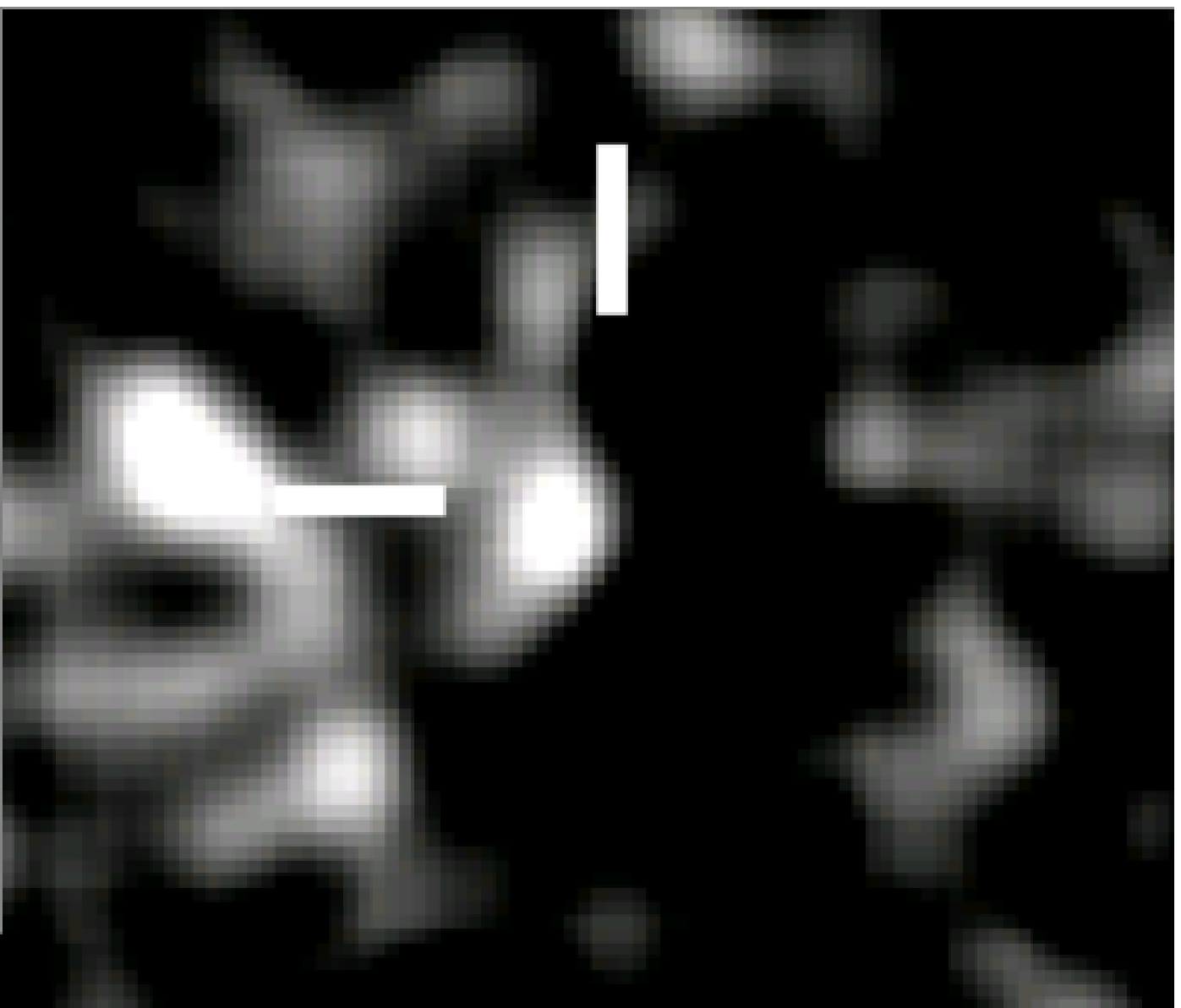}}
      \caption{Objects from \citet{Krivov} in the W3 (left) and W4 (right) bands. From top to bottom: CoRoT-8, HAT-P-5, and TrES-2 (KIC~11446443). We reject them as candidates because all are offset in the W4 band with respect to their shorter-wavelength counterparts. As a secondary reason, we note that CoRoT-8 is extended in W3.}\label{fig:krivov}
\end{figure*}

By performing an analogous analysis, \citet{Krivov} selected HAT-P-5, and CoRoT-8 as candidate sources with excesses in W4 and TrES-2 and XO-5 as candidates sources with excesses at W3 and W4. However, we do not confirm any of these as excess sources. For TrES-2 (KIC~11446443) we do not detect excess in W3 with high enough significance. The SDSS optical photometry coupled with the VOSA analysis confirms the lower effective temperature reported by \citet{odonovan2006} as compared with the higher best-fit temperature found by \citet{Krivov}. That makes the observed W3 flux not so different from the photospheric value ($\chi_{12}=0.9$) and therefore we conclude that the excess is not significant. For TrES-2, HAT-P-5, and CoRoT-8 we detect excess in W4 but direct inspection of the WISE images shows that the photocenter is offset with respect to shorter wavelengths (see Figure \ref{fig:krivov}). In the case of XO-5 we do not detect excesses at any band: For W4 we set a more stringent cut-off of $\chi_{22} >$ 2 and for W3 we again use the SDSS optical photometry, which results in no excess being predicted in the IR. In general, we note that the fits presented by  \citet{Krivov} yield systematically higher best-fit effective temperatures than published values and poorer fits to the stellar models than the ones we obtained with our Tycho/SDSS photometry and VOSA. Unfortunately, those authors do not report the best-fit extinction values nor explain whether they computed synthetic photometry by convolving the stellar models with the filter passbands. In summary, we cannot confirm any of the objects mentioned in \citet{Krivov} as having excesses in the WISE bands.

\begin{table*}
  \begin{center}
    \begin{tabular}{l c c c c c c c c c c c c c c c}
      \hline\rule{0mm}{3mm} Name & ${\rm SpT}$  &
      ${\rm a_{planet}[AU]}$ & ${\rm d_{star}[pc]}$ &  ${\rm F_{12}/F^*_{12}}$
      & $\chi_{12}$ & ${\rm F_{22}/F^*_{22}}$ & $\chi_{22}$ & ${\rm T_{dust}[K]}$ &  ${\rm R_{dust}[AU]}$ & $f={\rm L_{dust}/L_*[10^{-5}]}$
      \vspace{1mm}
      \\\hline
      \multicolumn{11}{c}{Quality flag (1) - Single isolated clean PSF}
      \\\hline \object{KIC~2853093 / KOI~1099} & G6 &  0.573 & 1300 & 7.1$\pm$1.0 & 6.1 & 58.7$\pm$28.2 & 2.2 & 200 & \,\,\,\,1.0  & \,\,\,\,6.5\\
       \object{KIC~4918309 / KOI~1582} & G9 & 0.626 & 1020 & 2.7$\pm$0.7 & 2.2 & $<46.4$ & \ldots & $>150$  & $<1.9$  & $<5.4$ \\
       \object{KIC~6422367 / KOI~559} & K0 & 0.052 & 660 & 2.5$\pm$0.4 & 4.0 & $<40.2$ & \ldots & $>150$   & $<2.7$  & $<4.7$  \\
      \object{KIC~6924203 / KOI~1370} & G8 & 0.071 & 910 & 2.1$\pm$0.6 & 2.2 & $<39.7$ & \ldots & $>150$  & $<2.4$  & $<3.7$  \\
      \object{KIC~9008220 / KOI~466} & G1 & 0.087 & 1100 & 2.1$\pm$0.5 & 2.2 & $<31.8$ & \ldots & $>150$  & $<2.1$  & $<2.9$  \\
      \object{KIC~9703198 / KOI~469} & G0 & 0.095 & 1200 & 3.2$\pm$0.5 & 4.2 & $<30.7$ & \ldots & $>175$  &  $<2.6$  & $<2.7$ \\
       \object{KIC~10526549 / KOI~746} & K3 & 0.080 & 570 & 1.7$\pm$0.4 & 2.1 & $<30.1$ & \ldots  & $>125$   & $<2.6$  & $<5.2$ \\
      \hline
      \multicolumn{11}{c}{Quality flag (2) - Possible source confusion or weak detection at W3 or W4}
      \\\hline \object{WASP-46} & G6 & 0.024 & 520 & 1.5$\pm$0.3 & 3.3 & $<6.7$ & \ldots & $>175$  & $<2.2$  & $<0.4$ \\
     \object{KIC~3547091 / KOI~1177} & G9 & 0.043 & 1080 & 2.9$\pm$0.9 & 2.2 & $<55.5$ & \ldots & $>150$  & $<2.0$  & $<5.6$ \\
     \object{KIC~3732821 / KOI~1207} & K1 & 0.111 & 720  &  2.4$\pm$0.6 & 2.6 & $<56.5$ & \ldots & $>125$  & $<3.6$  & $<10.0$ \\
      \object{KIC~4545187 /  KOI~223} & K0 & 0.041, 0.226 & 600 &  1.9$\pm$0.3 & 2.6 & $<20.9$ & \ldots & $>150$  & $<2.0$  & $<2.8$ \\
      \object{KIC~6665695 /  KOI~561} & K1 & 0.058 & 430 & 1.1$\pm$0.2 & 0.8 & 12.8$\pm$5.7 & 2.2 & 100  & \,\,\,\,4.8  & \,\,\,\,1.9 \\
      \object{KIC~8414716 /  KOI~910} & K1 & 0.057 & 850 & 2.5$\pm$0.7 & 2.2 & $<79.8$ & \ldots & $>125$  & $<2.0$   & $<14.8$  \\
      \hline\rule{0mm}{3.0mm}
      
    \end{tabular}
    \caption{Parameters of the selected objects. The candidates are
      separated in two groups based on the visual inspection
      process. $\rm SpT$ and $\rm a_{planet}$ from
      \citet{Borucki2011,WASP46}. For WASP-46, we use $\rm T_{eff}$= 5750 K and A$_V$=0. Stellar distances are calculated
      deriving the absolute $J$ magnitude from the $\rm T_{eff}$ or
      the $J-K$ color.$\rm T_{dust}$ is the temperature of a blackbody
      fitted to the W3 and W4 bands. The disk radius $\rm R_{dust}$ is
      derived assuming large dust grains. $f=\rm L_{dust}/L_*$ is
      derived from the fitted blackbody.  KIC~4545187 is a multiple planet system.}\label{tab:param}
  \end{center}
\end{table*}

\begin{figure*}[H!]
  \begin{center} 
    \subfigure{\includegraphics[width=0.3\hsize]{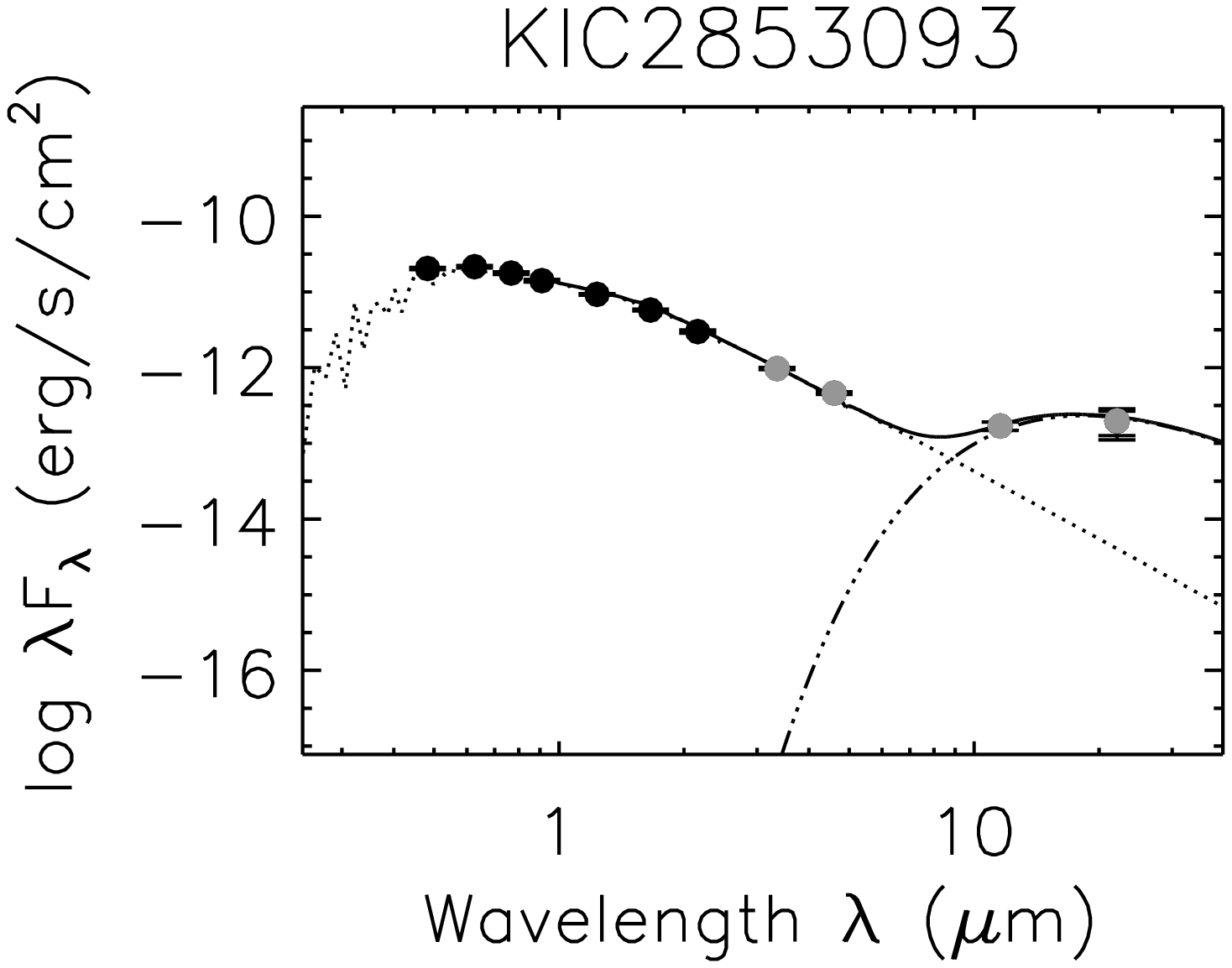}}
    \subfigure{\includegraphics[width=0.3\hsize]{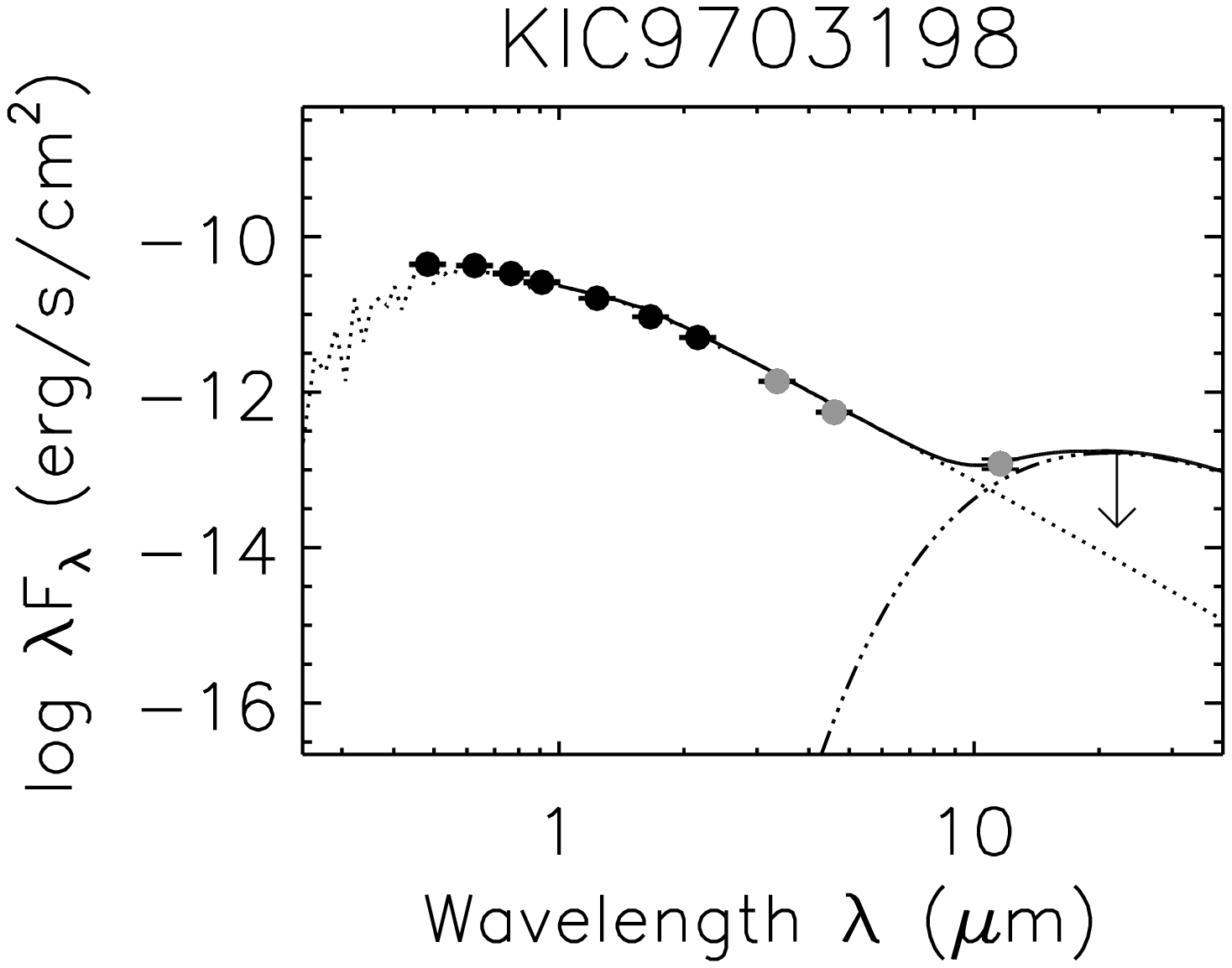}}
    \subfigure{\includegraphics[width=0.3\hsize]{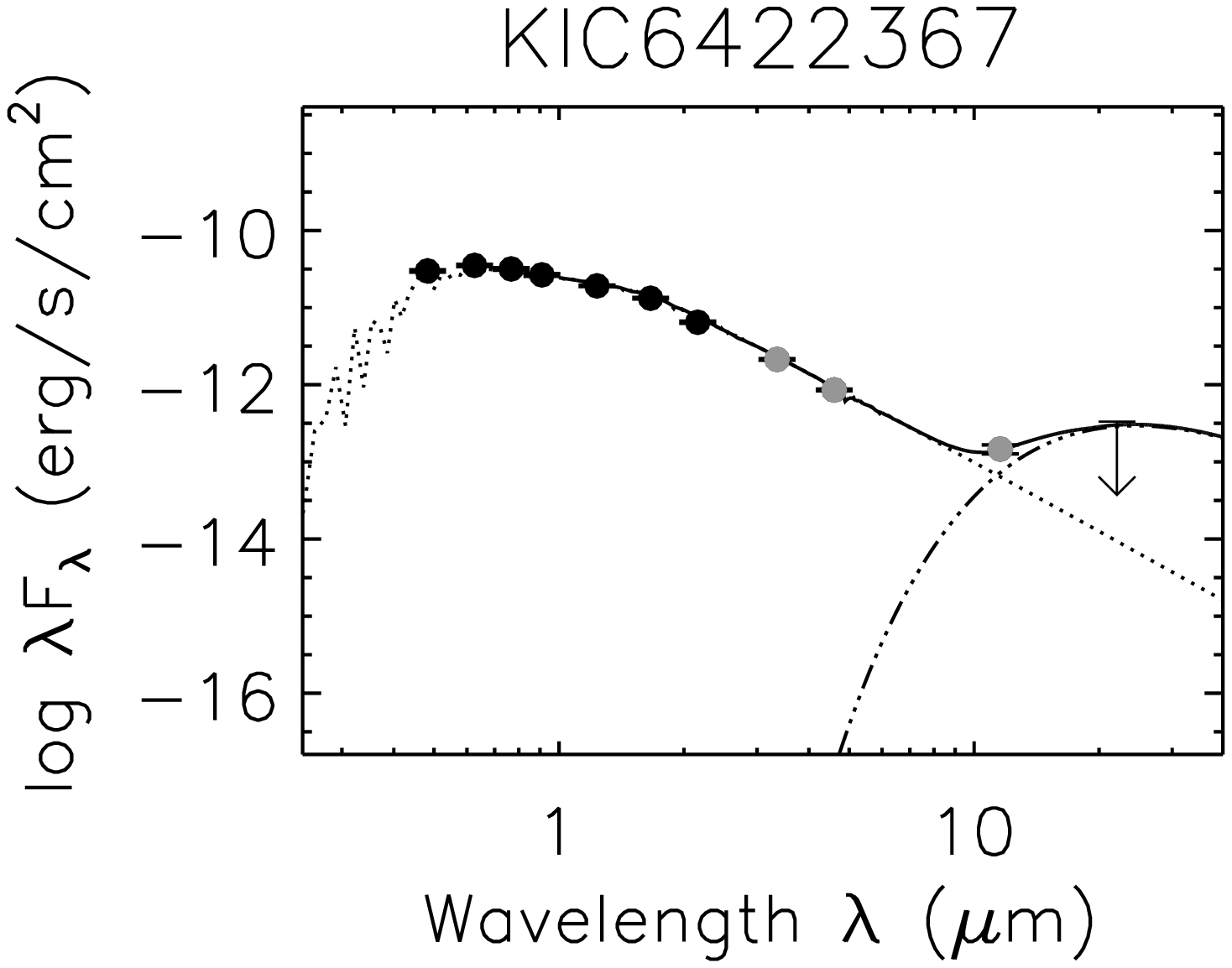}} \\
    \subfigure{\includegraphics[width=0.3\hsize]{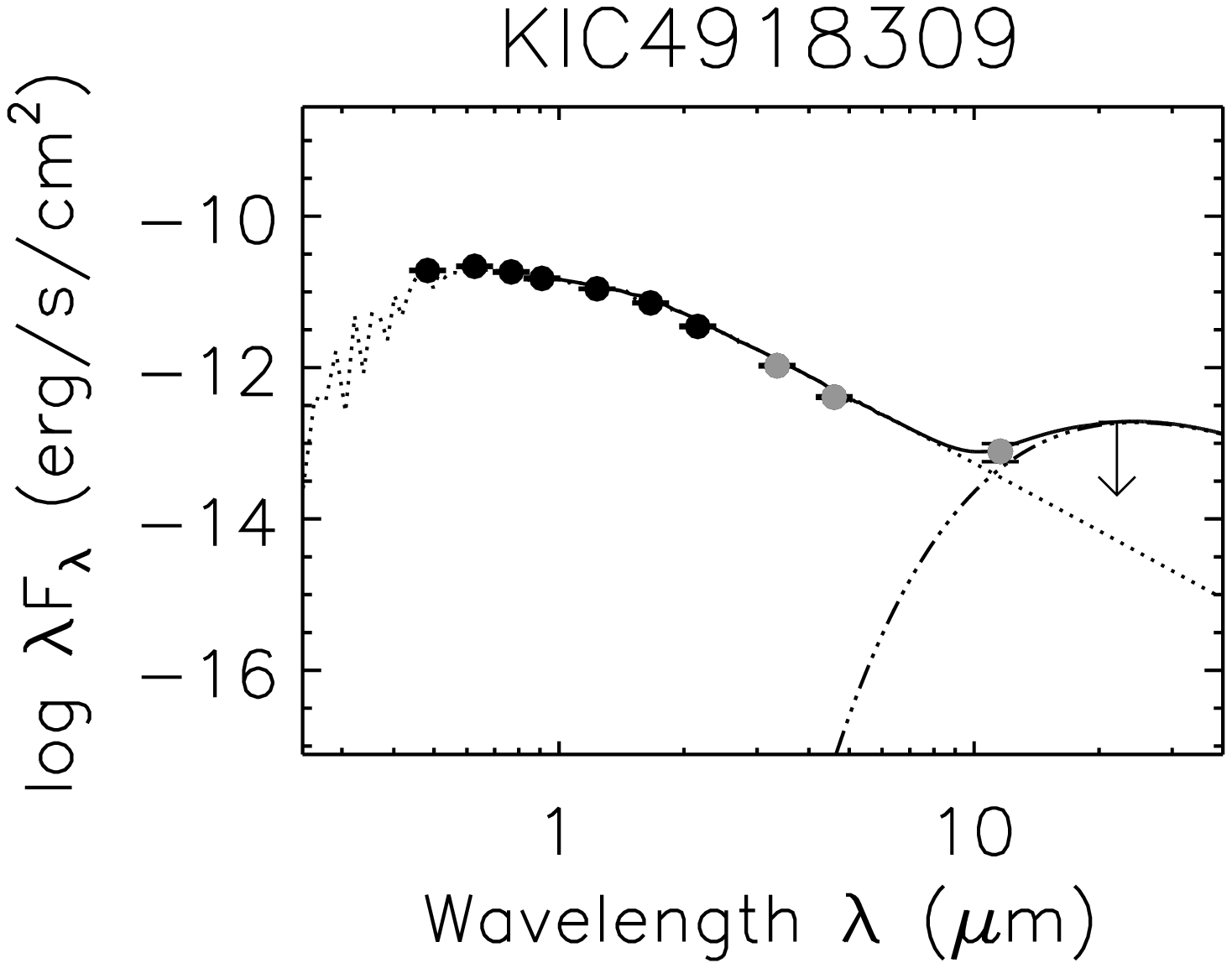}} 
    \subfigure{\includegraphics[width=0.3\hsize]{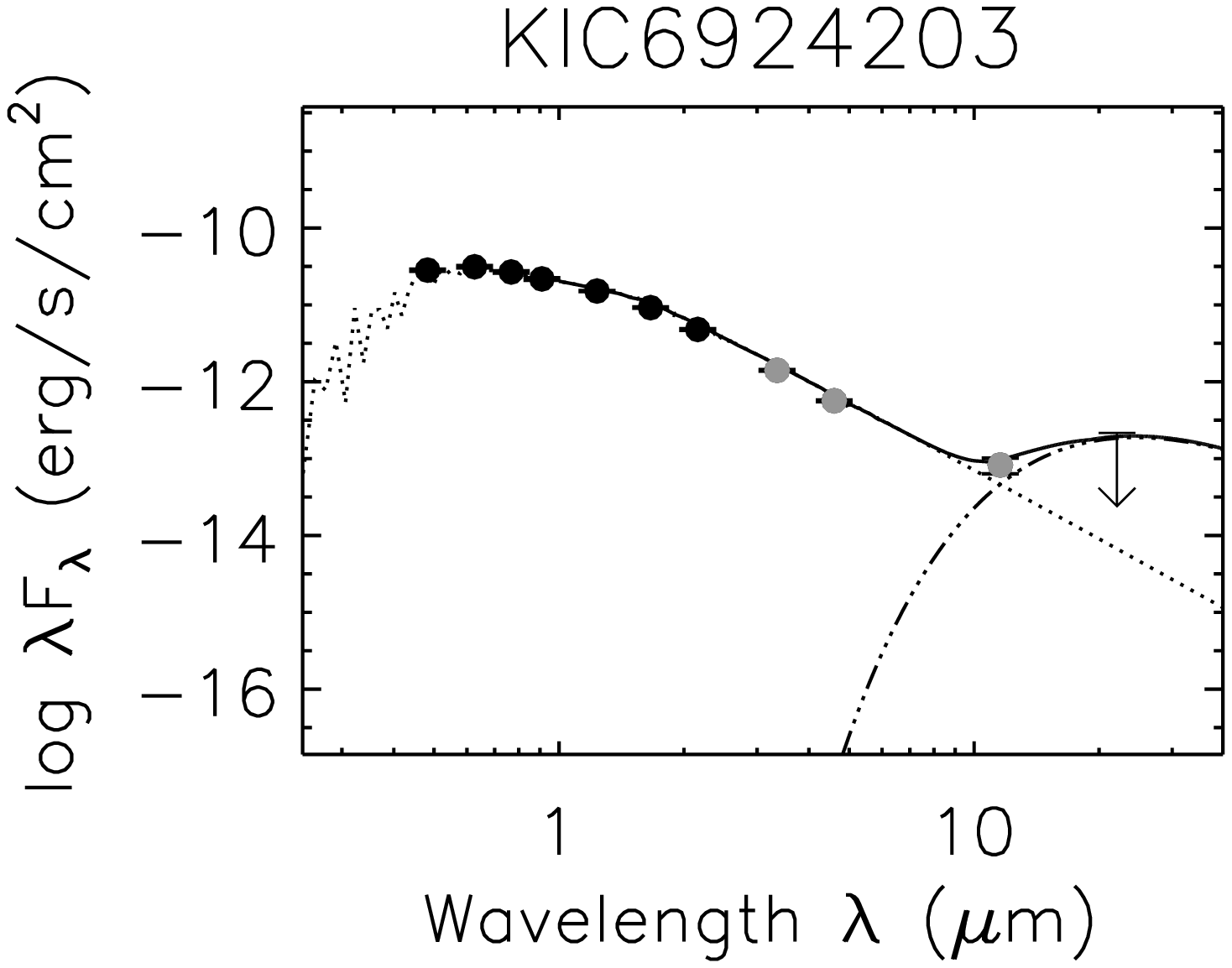}}
    \subfigure{\includegraphics[width=0.3\hsize]{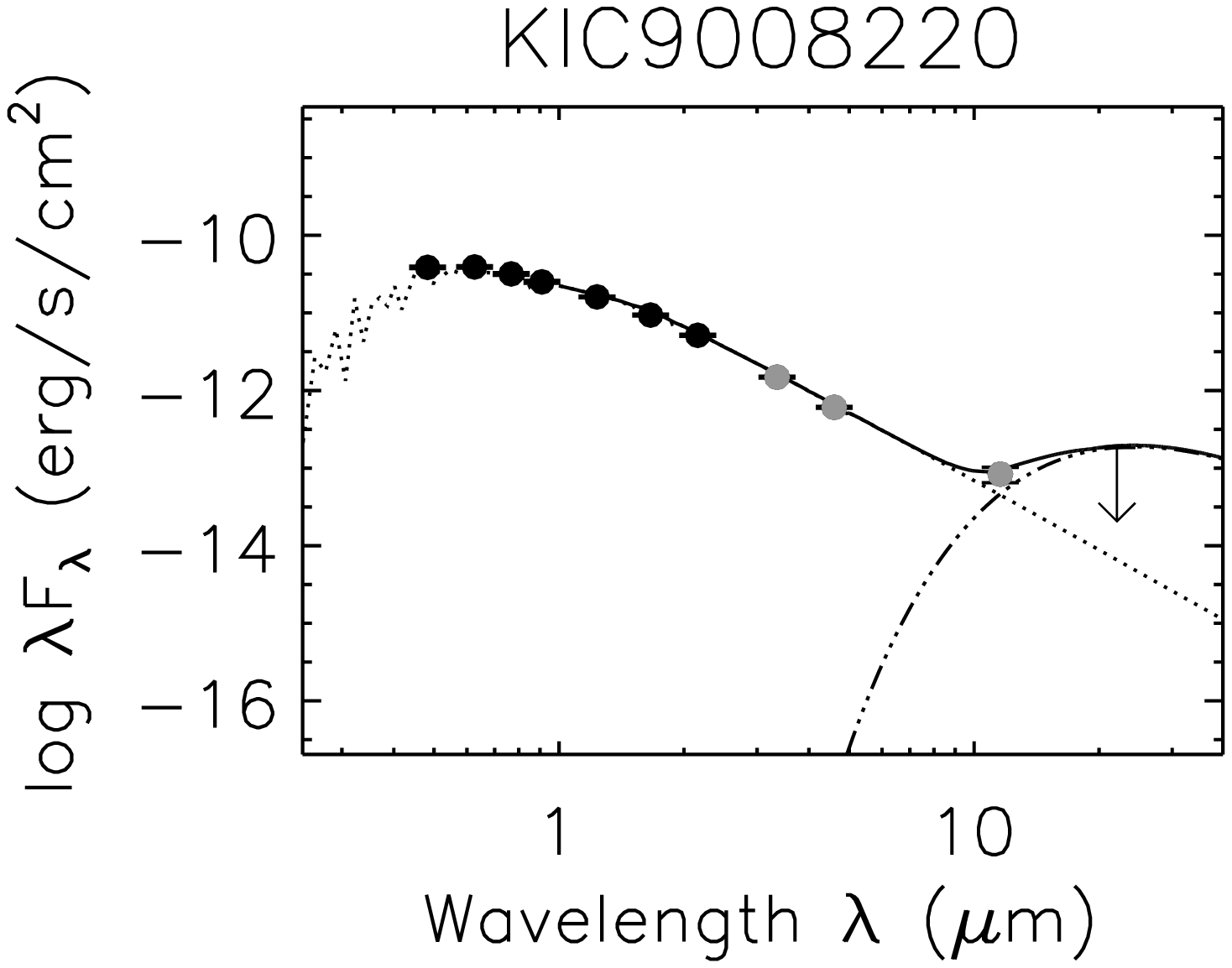}} \\
    \subfigure{\includegraphics[width=0.3\hsize]{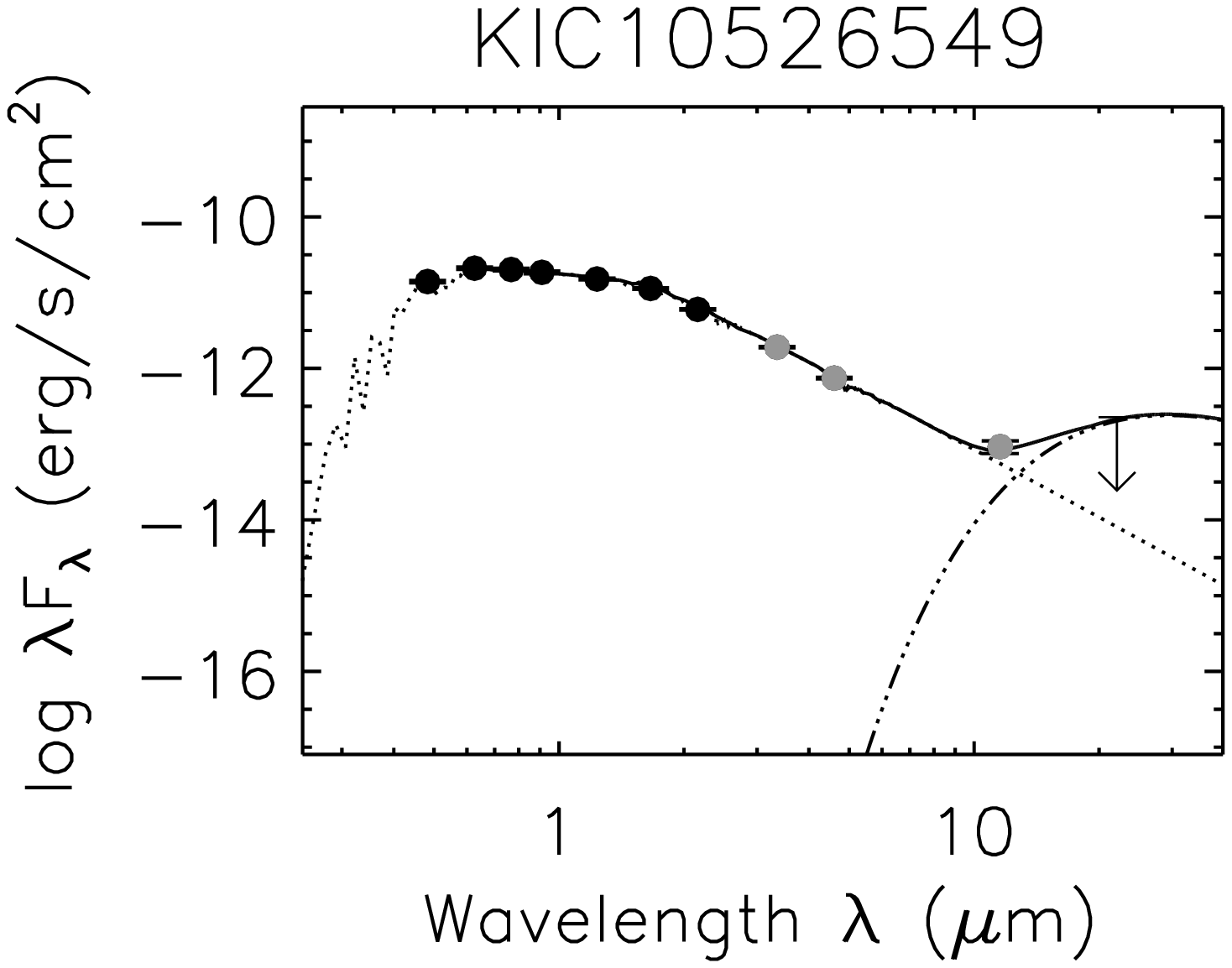}}
    \subfigure{\includegraphics[width=0.3\hsize]{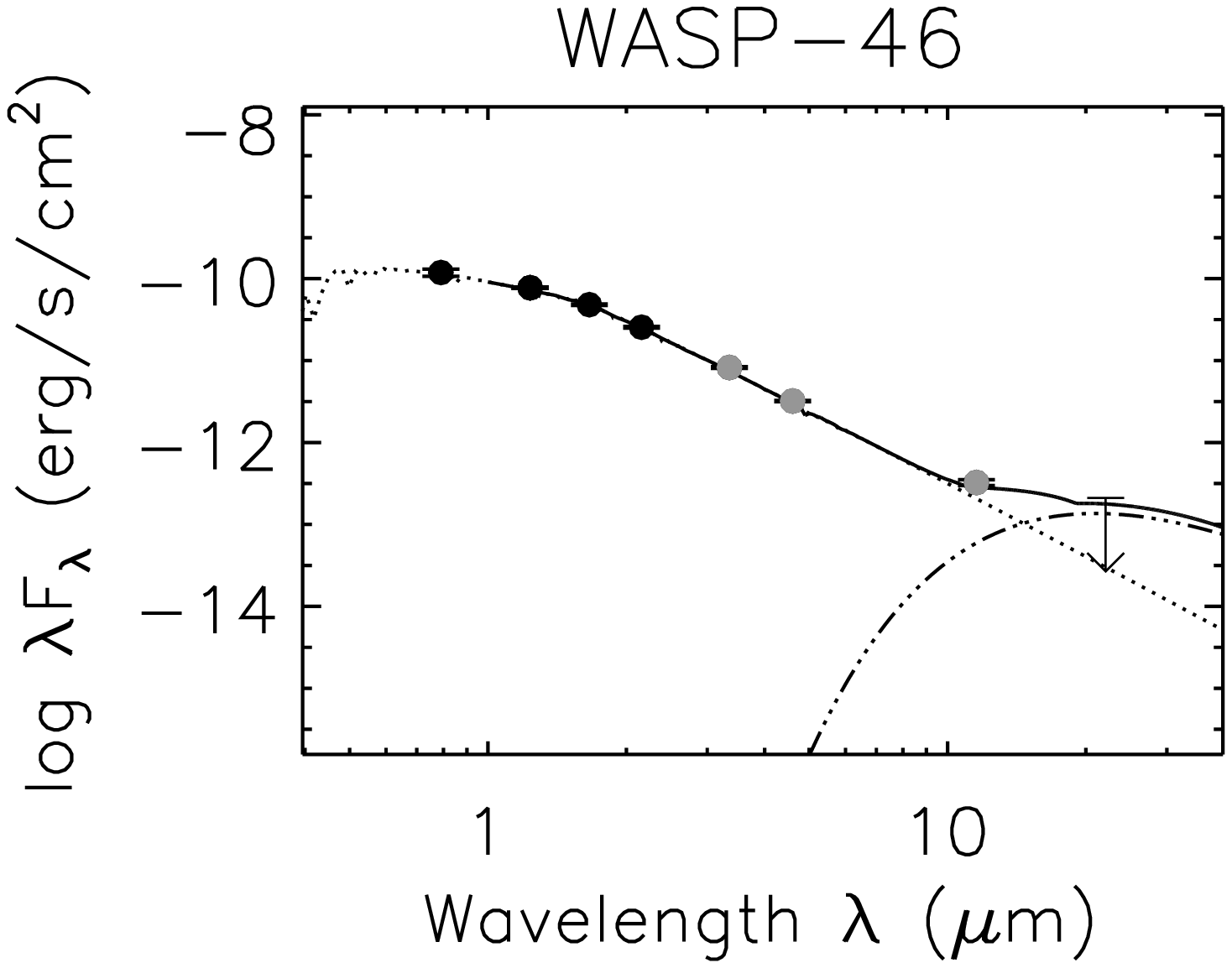}}
    \subfigure{\includegraphics[width=0.3\hsize]{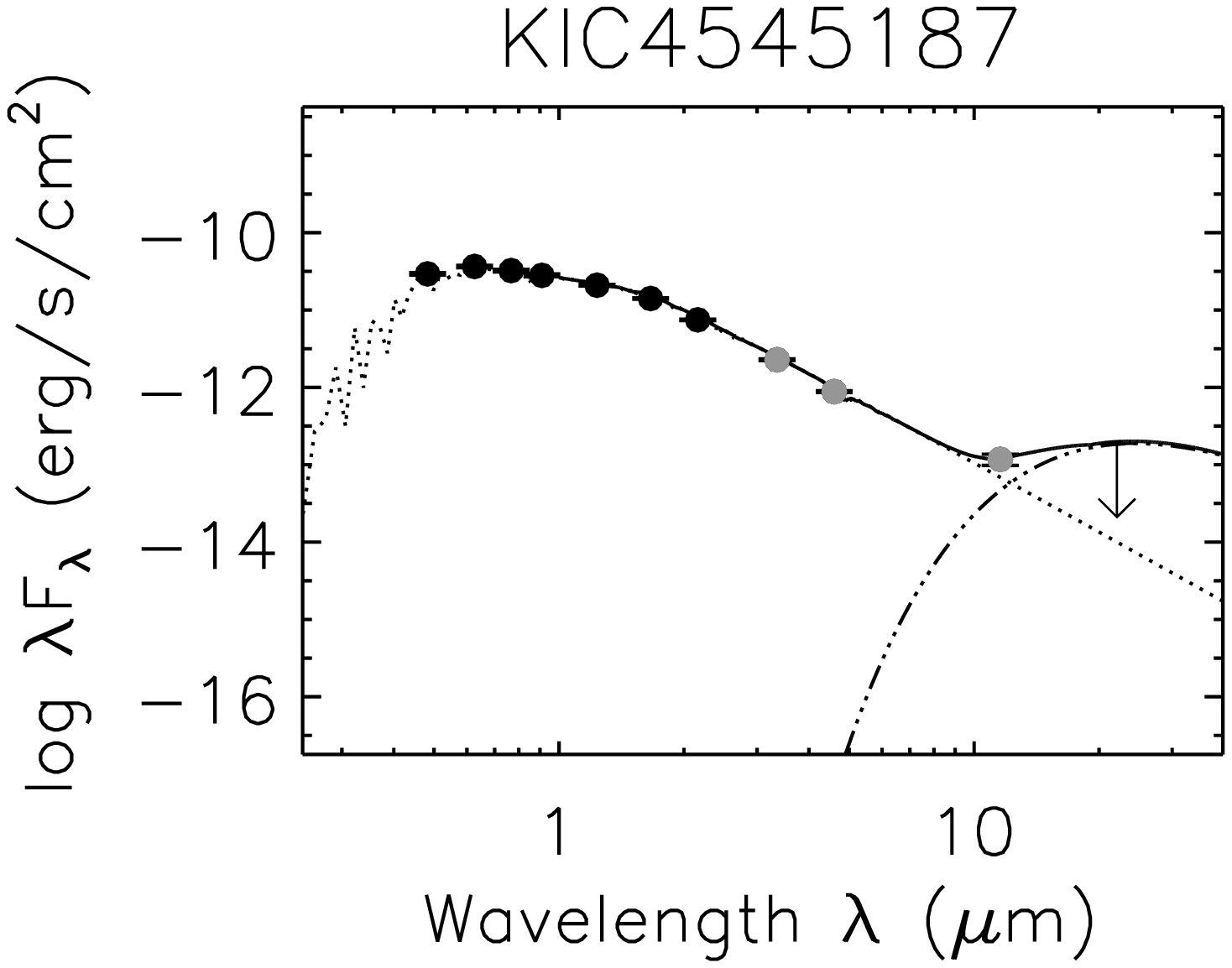}} \\
    \subfigure{\includegraphics[width=0.3\hsize]{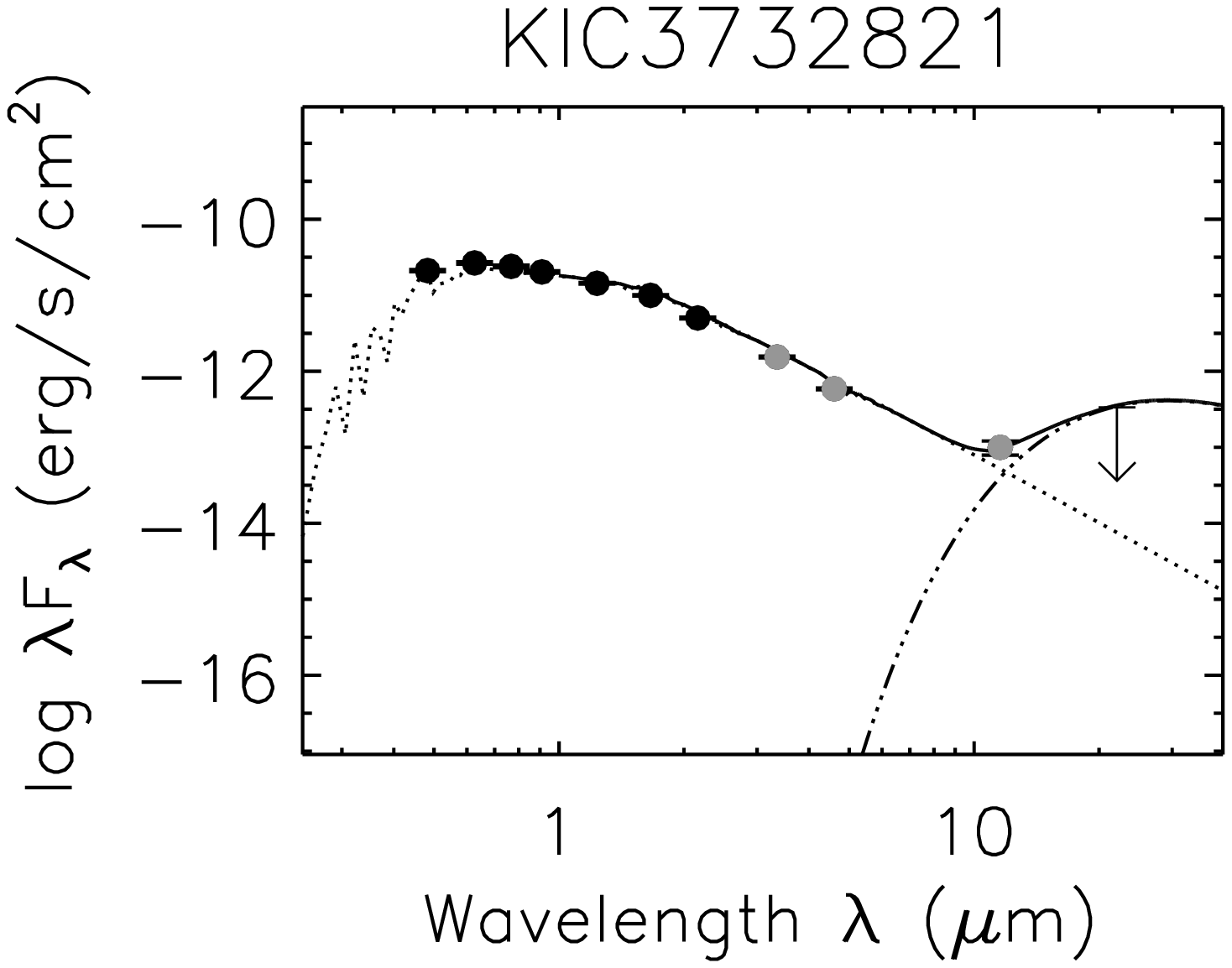}}
    \subfigure{\includegraphics[width=0.3\hsize]{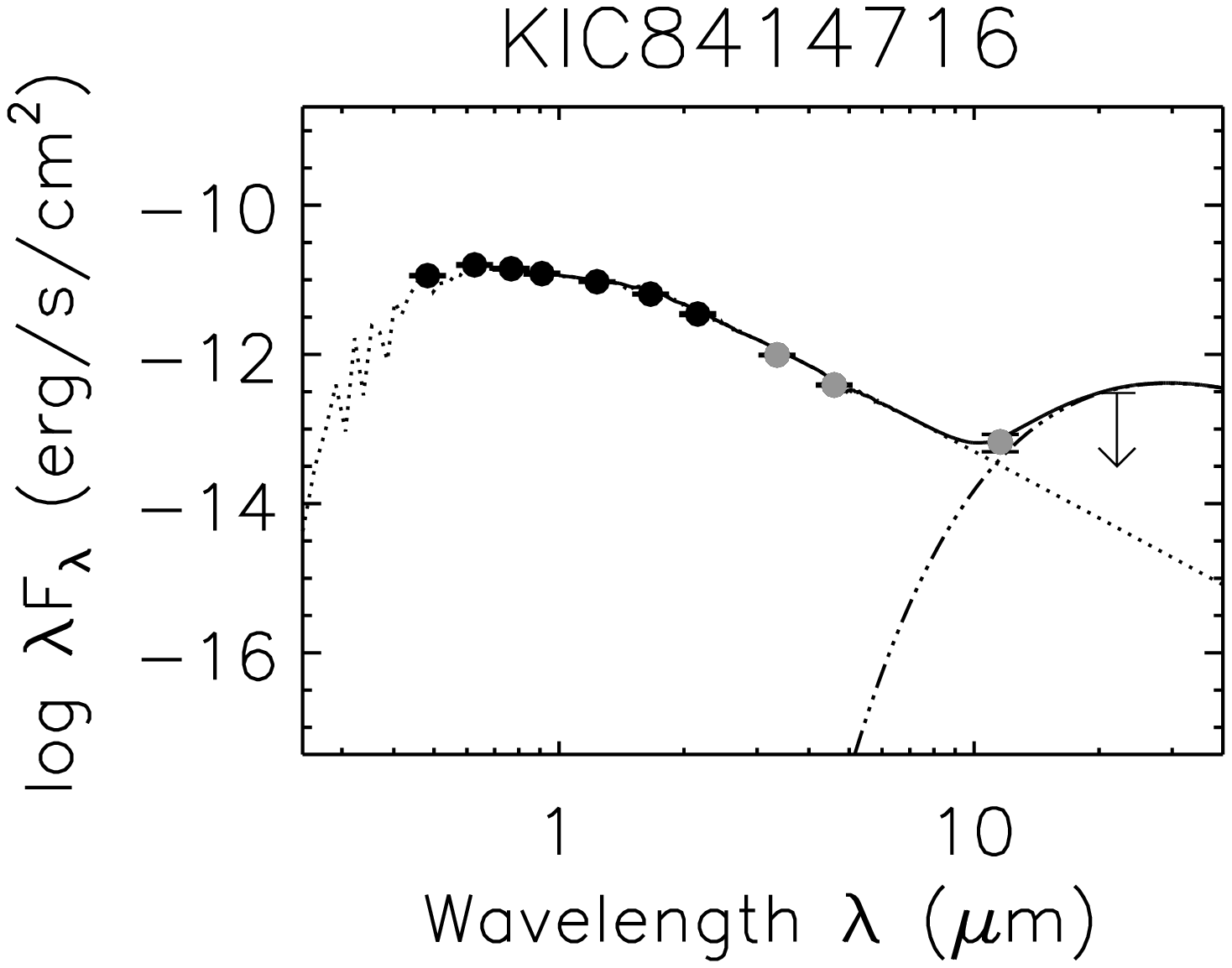}}
    \subfigure{\includegraphics[width=0.3\hsize]{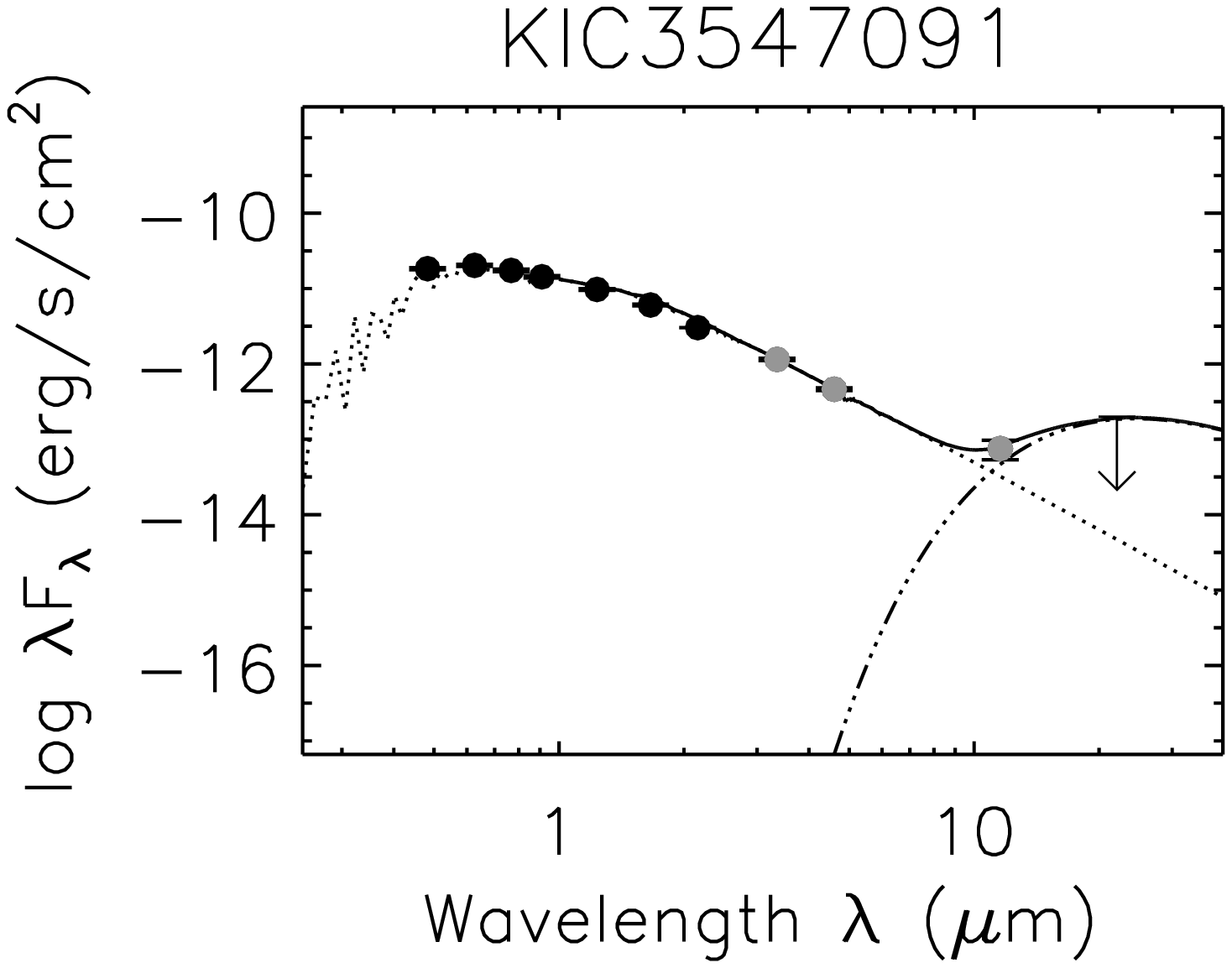}} \\
    \subfigure{\includegraphics[width=0.3\hsize]{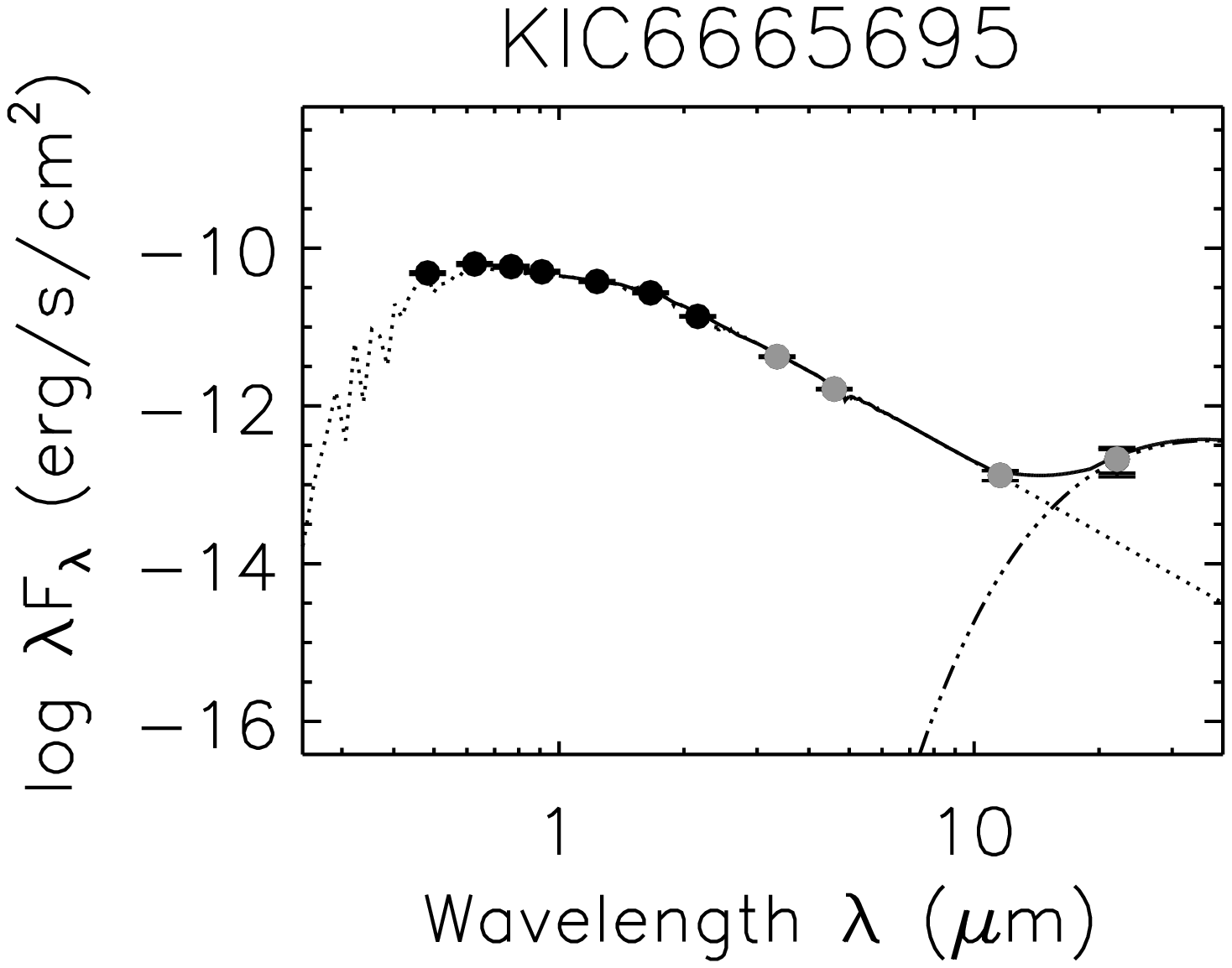}}
    \caption{SEDs of the selected candidates. The fitted dereddened
      photometry (solid line), optical and 2MASS (black dots) and WISE
      (grey dots) are represented. A photosphere (dotted line) is
      fitted to the 2MASS and W1 and W2 bands and the thin line
      represents the total of stellar plus excess emissions. Arrows
      represent W4 upper limits.}\label{fig:SEDS}
  \end{center}
\end{figure*}


\section{POSSIBLE EXPLANATIONS FOR THE INFRARED EXCESS}\label{sec:explanations}

As shown in Fig.~\ref{fig:SEDS}, the IR portion of the SEDs cannot be
fitted by a pure photosphere. We characterize the excess by the color temperature obtained when fitting a single blackbody function that reproduces the excesses at W3 and the detections or upper limits in W4 (see Tab.~\ref{tab:param}). Lack of substantial excesses at W2 or shorter wavelengths implies upper limits to the dust color temperatures of $\leq$ 500 K. We explicitly refrain from including in our analysis the potential excesses found in two sources at W1 and W2 bands until the final WISE catalog has been released and there is a better understanding of the uncertainties at those bands for these sources.

The observed IR excess could be due to: (1) The contribution of a cool companion; (2) The random alignment with a foreground object; (3) A background galaxy; (4) Interaction between the host star and an interstellar medium (ISM) cloud; (5) A Debris Disk.  We now examine each of these possibilities.

To reproduce the color temperature, a cool companion would need to be a Y-dwarf like the recently discovered WISE 1541-2250 \citep{Kirkpatrick2011}. With an effective temperature of 300-450 K, such an object would produce excess in the W3 and W4 bands, as observed. Using WISE 1541-2250 as template, its W4 magnitude at 430 pc  (the distance to the closest candidate) would be  $>$19 mags. From the WISE PRC, the 5$\sigma$ limit at W4 is $\sim$8 mags\footnote{http://wise2.ipac.caltech.edu/docs/release/prelim/}. Therefore, any such cool companion would have to be anomalously bright to contribute to the excess in the IR if bound or happen to be 
co-aligned with the stars at a distance from the Sun smaller than 45 pc, both of which scenarios are relatively unlikely.

WISE 1541-2250 has an observed W3 magnitude of 12.1 mags, implying $Log(\lambda F_\lambda) \sim -13$ dex at 12 $\mu$m. As shown in Fig. \ref{fig:SEDS} this is comparable to the excess fluxes we detect. The work by \citet{Kirkpatrick2011} used all-sky WISE data to search for Y-dwarfs and found 6 confirmed objects. Assuming they are uniformly distributed in the sky, the probability of finding one within 100 squared-arcseconds of one of our targets is a few parts in a million. Therefore we consider this explanation to be very unlikely as well.

In principle, a background galaxy could be responsible for the IR excess. We searched the NASA Extragalactic Database (NED) for any galaxy within 10" of the candidates and found none. Furthermore, {\it Spitzer's} MIPS galaxy counts at 24 $\mu$m indicate that we can expect $\sim 5\times10^{-5}$ infrared luminous galaxies  brighter than 5 mJy at 24 $\mu$m on average within 10" of any of our sources \citep{Papo2004}. As in the previous cases, this scenario is found to be too improbable to explain the IR excesses.
 
The interaction with an ISM cloud is a more realistic possibility.  Given that the candidates are
located beyond the Local Bubble, the IR excess may be due to the
so-called Pleiades Effect, the random alignment between a clump of
interstellar medium and a star \citep{Kalas2002}. The role of this effect is difficult to estimate a-priori, as dust maps do not have the required spatial resolution. \citet{Kalas2002} have shown that the ISM dust would emit mostly at longer wavelengths, although populations of very small particles may result in shorter wavelength emission. Mid-IR spectroscopic observations will clarify the particle size of the dust responsible for the excess, and longer wavelength observations may help rule out the role of ISM altogether. With the data at hand, it is not possible to rule out this phenomenon. 

\section{DEBRIS DISKS IN THE TRANSITING PLANET SAMPLE}

By assuming that the IR excess is due to the presence of a debris disk made up of large particles, we derive distances between the star and the dust  using its conection with the luminosity of the host star given by equation 3 in \citep{Wyatt2008}. The distances we obtain are comparable to the planetary semi-major axis  (see Tab.~\ref{tab:param}).

These distance estimates depend critically on the assumed particle size. For 0.2\,$\mu$m grains, the typical size of the radiation pressure blow-out limit for a G0 star \citep{Krivov}, a dust ring 8 AU in radius would have a temperature of $\sim$ 150\,K,
assuming typical silicate emissivities \citep[see][and references
  therein]{Beichman2006}. A better handle on the dust distance can then be obtained by modeling mid-IR spectra of the dust, which will provide estimates of typical grain sizes.
 
About 2\% of the 546 objects considered here show IR excess consistent with the presence of a debris disk. We note however, that the sample is biased by the sensitivity levels of the WISE spacecraft, and therefore this excess fraction cannot be compared to the 4\% figure found by \citet{Trilling2008} at
$24\mu$m in a volume-limited sample. Furthermore, in most of the objects from that work the $24\mu$m
excess represents the short wavelength side of the emission, which
peaks at longer wavelengths. In contrast, for the candidates presented
here, the excess at 12\,$\mu$m plus the detection or the upper limit
at 22\,$\mu$m, allow us to definitely conclude that we are observing
warm dust.

\begin{figure*}
  \begin{center} 
    \vspace{5pt}
    \includegraphics[width=0.8\hsize]{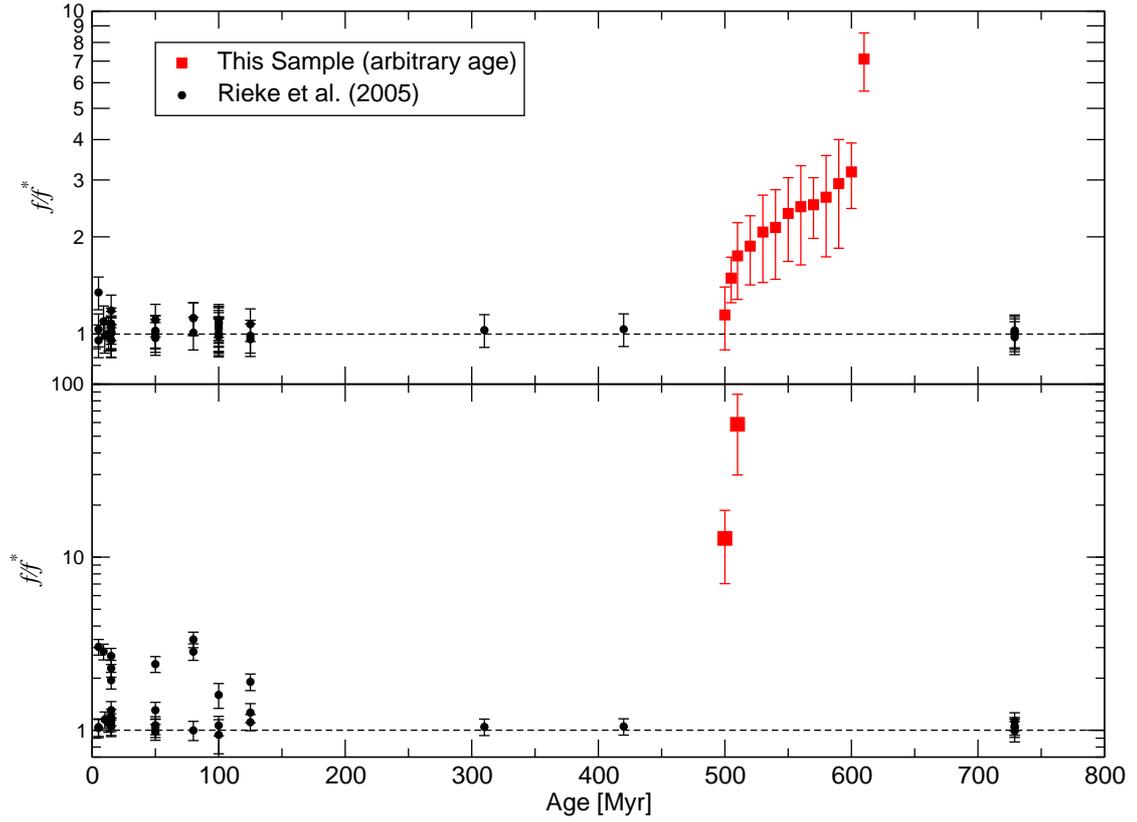}
        \caption{Excess as a function of time. The top plot shows the evolution of F$_{12}$/F$_*$, the observed WISE flux as a fraction of the predicted stellar flux,  as a function of time. The bottom plot shows the analogous quantity for 22 $\mu$m. The stellar sample is based on the WISE measurements of the \citet{Rieke2005} sample. The red points show the position of the candidates we present here. The ages of the KIC sample are generally unknown. We have arbitrarily assumed ages between 500 and 600 Myrs. To facilitate the comparison we have also set WASP-46 in this range.}\label{fig:rieke_wise}
  \end{center}
\end{figure*}

The existence of warm IR excess is correlated with age, appearing at a greater frequency for young stars \citep{Rieke2005,Siegler2007,Trilling2008}. However, the KIC sample was explicitly selected to contain only low-activity, large surface gravity (dwarf and sub-giant) stars \citep{KICselection}. WASP-46 has an estimated age of 1.4 Gyr \citep{WASP46}.  At these ages, most stars do not present excess at 24 $\mu$m \citep[e.g.][]{Siegler2007}. 
In order to present a preliminary comparison of the WISE results presented here with the literature on debris disk decay, we performed a search for WISE counterparts to the Spitzer 24$\mu$m objects from  \citet{Rieke2005}. Following the same procedure mentioned above for the transiting planet sample, we obtained excesses at 12 and 22 $\mu$m for the  \citet{Rieke2005} targets. As shown in Figure \ref{fig:rieke_wise} the excesses for the WISE targets are well above the expected excesses due to debris disks of comparable ages. 

We note here that this is a preliminary comparison, based on the small sample of Spitzer objects with excesses at the WISE wavelengths and the photometry from the WISE PRC. However, it is remarkable that we should find such large excesses in the transiting planet sample. The objects presented here seem to be analogous, in terms of F$_\lambda$/F$_*$, to objects like $\eta$ Corvi \citep{Lisse2011}, HD~69830, BD+20307, and HD 23514 \citep{Wyatt2008}.

However, the values of $f$ as listed in Table \ref{tab:param} are small compared to those from other hot disks like $\eta$ Corvi ($f\sim 5 \times 10^{-4}$) or  HD~69830 ($f\sim 2 \times 10^{-4}$). \citet{Wyatt2007b} provide an estimate for the maximum fractional excess $f$ that one could expect as a function of age (see their equation 21).  For a 1 Gyr-old star, this number varies between $10^{-7}$ for dust at 1 AU, and $10^{-5}$ for dust at 10 AU. Our targets have $f$-values comparable to the later number, so it is not impossible that we are seeing dust produced by very massive asteroid belts.

If the particles are relatively large, and the dust belts are closer to their central stars, we enter the realm of parameter space comparable to other warm disks. Two different origins for hot dust have been proposed: either it is due to a recent collision between two large planetesimals in an asteroid belt \citep{Song2005}, or it is derived from a planetesimal belt further from the star where the mass required to sustain the inferred mass-loss rate can survive for long periods of time \citep{Absil2006,Wyatt2007b}.  

Without further data we cannot weight on this debate. However, we note that the presence of large planetesimals close to the star or the existence of a dynamical instabilities analogous to the Late Heavy Bombardment in our Solar System may be the consequence of the presence of the transiting planets. An explanation for the excess in HD~69830 may be in its three Neptune-mass planets \citep{Lovis2006}. Other studies have failed to find any correlation between the presence of planets and the existence of cold debris disks \cite{Kospal2009,Bryden2009},  but our results suggest that the presence of transiting planets in a warm debris system may cause the stirring of the disk, resulting in increased dust production detectable in the WISE bands in these systems. Whether this is because the planets perturb asteroidal disks, favor the orbital decay of planetesimals from farther away, or perturb far-away planetesimals is unclear. It is also unclear whether this happens in non-planet systems.

In conclusion, longer-wavelength photometric and spectroscopic observations are crucial for the further characterization of these systems. The Kepler and WISE samples are veritable goldmines in the study of planet formation and we have just begun scratching the surface. 

\section{CONCLUSIONS}

We present a new sample of 13 transiting planets systems with
significant ($\chi_\lambda>2$) IR excess at 12 and/or 22\,$\mu$m
obtained matching the Kepler Input Catalog and a sample of previously
known transiting planets hosts with the WISE Preliminary 
Release Catalog (PRC) in the overlapping areas. We describe in
detail the quality control procedure applied to identify trustworthy
excesses in the sources, which includes estimates of significance of
the excesses and inspection of the individual WISE images. We further 
conclude that the IR excesses identified by \citet{Krivov} may be spurious.
This process will be repeated once the final WISE all-sky point source
catalog is released and covers the whole Kepler field.

While we cannot rule out that this excess is due to ISM clumps illuminated by the stars, we assume that they are due to debris 
disks. Assuming large dust particles, the SEDs are consistent with inner disk radii 
comparable to the planets semi-major axes. The infrared excess as fraction 
of stellar flux is larger than expected for old field stars, and we suggest that the  
planets are stirring the disks.

\begin{acknowledgements}

  We thank the referee for valuable comments that helped improving the
  contents of this paper. This work has been possible thanks to the
  support from the ESA Trainee and ESAC Space Science Faculty and of
  the Herschel Science Centre. This publication is based on
  observations made with the Kepler Spacecraft. Funding for this
  mission is provided by National Aeronautics and Space
  Administration's Science Mission Directorate (NASA). This study also
  makes use of VOSA, developed under the Spanish Virtual Observatory
  project supported from the Spanish MICINN through grant
  AyA2008-02156; data products from the Wide-field Infrared Survey
  Explorer, a joint project of the University of California, Los
  Angeles, and the Jet Propulsion Laboratory (JPL) / California
  Institute of Technology (Caltech); the NASA Infrared Processing and
  Analysis Center (IPAC) Science Archive and the NASA/IPAC/NExScI Star
  and Exoplanet Database, operated by JPL/Caltech, and funded by NASA;
  the SIMBAD database and the Vizier service, operated at the Centre
  de Données astronomiques de Strasbourg, France; the data products
  from the Two Micron All Sky Survey (2MASS), a joint project of the
  University of Massachusetts and IPAC/Caltech, funded by NASA and the
  National Science Foundation; the Multimission Archive at the Space
  Telescope Science Institute (MAST). STScI is operated by the
  Association of Universities for Research in Astronomy, Inc., under
  NASA contract NAS5-26555. Support for MAST for non-HST data is
  provided by the NASA Office of Space Science via grant NNX09AF08G
  and by other grants and contracts.
\end{acknowledgements}

\bibliographystyle{aa}
\bibliography{mybiblio}

\end{document}